\documentclass[aps,pra,reprint,showpacs]{revtex4-1}
\usepackage{amsmath}
\usepackage{SIunits}
\usepackage{graphicx}

\begin{document}
%
%
%
\newcommand{\ac}[0]{\ensuremath{\hat{a}_{\mathrm{c}}}}
\newcommand{\acmax}[0]{\ensuremath{\hat{a}_{\mathrm{c}}^{\mathrm{max}}}}
\newcommand{\adagc}[0]{\ensuremath{\hat{a}^{\dagger}_{\mathrm{c}}}}
\newcommand{\aR}[0]{\ensuremath{\hat{a}_{\mathrm{R}}}}
\newcommand{\aT}[0]{\ensuremath{\hat{a}_{\mathrm{T}}}}
\renewcommand{\b}[0]{\ensuremath{\hat{b}}}
\newcommand{\bdag}[0]{\ensuremath{\hat{b}^{\dagger}}}
\newcommand{\betaI}[0]{\ensuremath{\beta_\mathrm{I}}}
\newcommand{\betaR}[0]{\ensuremath{\beta_\mathrm{R}}}
\newcommand{\bra}[1]{\ensuremath{\left<#1\right|}}
\renewcommand{\c}[0]{\ensuremath{\hat{c}}}
\newcommand{\cdag}[0]{\ensuremath{\hat{c}^{\dagger}}}
\newcommand{\CorrMat}[0]{\ensuremath{\boldsymbol\gamma}}
\newcommand{\Deltacs}[0]{\ensuremath{\Delta_{\mathrm{cs}}}}
\newcommand{\Deltae}[0]{\ensuremath{\Delta_{\mathrm{e}}}}
\newcommand{\dens}[0]{\ensuremath{\hat{\rho}}}
\newcommand{\erfc}[0]{\ensuremath{\mathrm{erfc}}}
\newcommand{\gammap}[0]{\ensuremath{\gamma_{\mathrm{p}}}}
\newcommand{\gammapar}[0]{\ensuremath{\gamma_{\parallel}}}
\newcommand{\gammaperp}[0]{\ensuremath{\gamma_{\perp}}}
\newcommand{\gbar}[0]{\ensuremath{\bar{g}}}
\newcommand{\gens}[0]{\ensuremath{g_{\mathrm{ens}}}}
\renewcommand{\H}[0]{\ensuremath{\hat{H}}}
\newcommand{\Hext}[0]{\ensuremath{\hat{H}_{\mathrm{ext}}}}
\newcommand{\Hint}[0]{\ensuremath{\hat{H}_{\mathrm{I}}}}
\renewcommand{\Im}[0]{\ensuremath{\mathrm{Im}}}
\newcommand{\kappac}[0]{\ensuremath{\kappa_{\mathrm{c}}}}
\newcommand{\ket}[1]{\ensuremath{\left|#1\right>}}
\newcommand{\mat}[1]{\ensuremath{\mathbf{#1}}}
\newcommand{\mean}[1]{\ensuremath{\langle#1\rangle}}
\newcommand{\omegac}[0]{\ensuremath{\omega_{\mathrm{c}}}}
\newcommand{\omegas}[0]{\ensuremath{\omega_{\mathrm{s}}}}
\newcommand{\pauli}[0]{\ensuremath{\hat{\sigma}}}
\newcommand{\Pa}[0]{\ensuremath{\hat{P}_{\mathrm{c}}}}
\newcommand{\Pmax}[0]{\ensuremath{P_{\mathrm{max}}}}
\renewcommand{\Re}[0]{\ensuremath{\mathrm{Re}}}
\renewcommand{\S}[0]{\ensuremath{\hat{S}}}
\newcommand{\Sminuseff}[0]{\ensuremath{\hat{S}_-^{\mathrm{eff}}}}
\newcommand{\Sxeff}[0]{\ensuremath{\hat{S}_x^{\mathrm{eff}}}}
\newcommand{\Syeff}[0]{\ensuremath{\hat{S}_y^{\mathrm{eff}}}}
\newcommand{\tildeac}[0]{\ensuremath{\tilde{a}_{\mathrm{c}}}}
\newcommand{\tildepauli}[0]{\ensuremath{\tilde{\sigma}}}
\newcommand{\Tfocus}[0]{\ensuremath{T_{\mathrm{focus}}}}
\newcommand{\Tmem}[0]{\ensuremath{T_{\mathrm{mem}}}}
\newcommand{\Tswap}[0]{\ensuremath{T_{\mathrm{swap}}}}
\newcommand{\Var}[0]{\ensuremath{\mathrm{Var}}}
\renewcommand{\vec}[1]{\ensuremath{\mathbf{#1}}}
\newcommand{\Xa}[0]{\ensuremath{\hat{X}_{\mathrm{c}}}}

\title{Fundamental limitations in spin-ensemble quantum memories for
  cavity fields}

\author{Brian Julsgaard}
\email{brianj@phys.au.dk}

\author{Klaus M{\o}lmer}
\affiliation{Department of Physics and Astronomy, Aarhus University, Ny
  Munkegade 120, DK-8000 Aarhus C, Denmark.}


\date{\today}

\begin{abstract}
  Inhomogeneously broadened spin ensembles play an important role in
  present-day implementation of hybrid quantum processing
  architectures. When coupled to a resonator such an ensemble may
  serve as a multi-mode quantum memory for the resonator field, and by
  employing spin-refocusing techniques the quantum memory time can be
  extended to the coherence time of individual spins in the
  ensemble. In the present paper we investigate such a memory protocol
  capable of storing an unknown resonator-field state, and we examine
  separately the various constituents of the protocol: the storage and
  read-out part, the memory hold time with the spin ensemble and
  resonator field decoupled, and the parts employing spin refocusing
  techniques. Using both analytical and numerical methods we derive
  how the obtainable memory performance scales with various physical
  parameters.
\end{abstract}


\maketitle

\section{Introduction}
\label{sec:Introduction}
During the past decade various physical ensemble systems have been
utilized as quantum memories for propagating optical fields. The first
experimental demonstrations typically employed alkali vapors
\cite{Julsgaard.Nature.432.482(2004), Chaneliere.Nature.438.833(2005),
  Eisaman.Nature.438.837(2005), Choi.Nature.452.67(2008),
  Honda.PhysRevLett.100.093601(2008),
  Appel.PhysRevLett.100.093602(2008)}, while rare-earth-metal ions in
solids have later been used \cite{deRiedmatten.Nature.456.773(2008),
  Hedges.Nature.465.1052(2010)}. In the solid-state implementations
the transition frequencies are inhomogeneously broadened due to
variations in the local environment, which on the one hand allows
multi-mode performance but on the other hand presents a challenge: The
retrieval of the stored quantum information requires the dephasing
caused by the frequency inhomogeneity to be reversed. To this end,
atomic frequency-comb (AFC) techniques
\cite{Afzelius.PhysRevA.79.052329(2009)} and controlled reversible
inhomogeneous broadening (CRIB)
\cite{Moiseev.PhysRevLett.87.173601(2001),
  Nilsson.OptCommun.247.393(2005)} were employed in
Refs.~\cite{deRiedmatten.Nature.456.773(2008)} and
\cite{Hedges.Nature.465.1052(2010)}, respectively. In both examples,
effectively homogeneous subsets of the inhomogeneously broadened
ensemble were prepared by hole-burning techniques---thus sacrificing
optical depth of the material to achieve coherence. For classical light
pulses, these preparation steps can be avoided using certain
spin-refocusing techniques \cite{Carlson.OptLett.8.483(1983),
  Lin.OptLett.20.1658(1995)}, essentially based on the Hahn echo
\cite{Hahn.PhysRev.80.580(1950)}. However, these have been shown to be
inapplicable at the quantum level due to noise generated by
excited-state absorbers \cite{Ruggiero.PhysRevA.79.053851(2009),
  Sangouard.PhysRevA.81.062333(2010)}. Nonetheless, a simple scheme
using two $\pi$-pulses has recently been proposed which uses a
silencing mechanism to prevent emission from the excited-state
ensemble after the first $\pi$-pulse while allowing a faithful
read-out from the non-inverted ensemble after the second $\pi$-pulse
\cite{Damon.NewJPhys.13.093031(2011)}. This revival-of-silenced-echo
(ROSE) protocol plays an important role in the present paper.

The key feature of the above-mentioned ensemble approaches is the
collective dipole-moment enhancement, which enables a sufficiently
strong free-space light-matter interaction. This is in stark contrast
to cavity quantum electrodynamics (CQED) where the radiation field is
enhanced by a high-finesse cavity in order to interact efficiently
with a single matter particle
\cite{Kimble.PhysicaScripta.T76.127(1998)}. In an intermediate regime
proposals exist to use inhomogeneously broadened ensembles coupled to
cavities of moderate finesse as quantum memories for propagating
optical fields \cite{Afzelius.PhysRevA.82.022310(2010),
  Moiseev.PhysRevA.82.022311(2010)}. The enhancement by a cavity
ensures sufficient optical depth in the AFC and CRIB quantum memory
schemes mentioned above.

In the microwave regime, the strong-coupling regime has recently been
reached between superconducting co-planar waveguide resonators and, on
the one hand, ensembles of electronic spins
\cite{Kubo.PhysRevLett.105.140502(2010),
  Schuster.PhysRevLett.105.140501(2010),
  Amsuss.PhysRevLett.107.060502(2011)}, and on the other hand,
superconducting Joseph-junction qubits
\cite{Wallraff.Nature.431.162(2004)}. Due to the tunability of such
resonators it is possible to construct hybrid quantum systems with the
cavity acting as a ``quantum bus'' between a ``processor'' and a
``memory'' unit \cite{Imamoglu.PhysRevLett.102.083602(2009),
  Wesenberg.PhysRevLett.103.070502(2009),
  Kubo.PhysRevLett.107.220501(2011),
  Yang.PhysRevA.84.010301R(2011)}. While preselection of spectral
portions of the spin ensemble constitutes a means to mitigate the
effects of inhomogeneous broadening
\cite{Bensky.PhysRevA.86.012310(2012)}, the tunability of the cavity
also enables the implementation of the ROSE protocol
\cite{Julsgaard.PhysRevLett.110.250503(2013),
  Afzelius.NewJPhys.15.065008(2013)}. Noting that electron-spin
degrees of freedom may even be transfered to nuclear-magnetic degrees
of freedom \cite{Fuchs.NaturePhys.7.789(2011)}, the superconducting
CQED has brought back the modern concepts of spin-refocusing quantum
memories to their origin of ESR or NMR.

In Ref.~\cite{Julsgaard.PhysRevLett.110.250503(2013)} we described how
the ROSE protocol can be implemented between a tunable
microwave-resonator quantum bus and a spin-ensemble quantum memory,
and the feasibility of the protocol in its entirety was assessed for
the special case using nitrogen-vacancy centers in diamond as the
memory unit. The present paper is targeted on isolating and assessing
the effect of the individual constituents of the protocol; the
spin-cavity transfer mechanism, the silencing mechanism, the
$\pi$-pulse process, and the role of decoherence. Hence, our work is
intended to present a firm base for developing and optimizing specific
quantum-memory protocols for intra-cavity fields using spin
ensembles.

The paper is arranged as follows: In Sec.~\ref{sec:Eq_of_motion} the
basic equations and assumptions are presented for the spin ensemble
system coupled to a cavity, and in Sec.~\ref{sec:Basic_idea} it is
briefly reviewed how this system enables a quantum memory
protocol. The actual examination of the quantum memory performance
begins in Sec.~\ref{sec:Spin-cavity-swap} with an account for the
storage and read-out part of the protocol, in
Sec.~\ref{sec:spin-cavity-decoupling} the decoupling of the spin
ensemble from the cavity is examined, and in
Sec.~\ref{sec:extern-appl-inversion} the effect of refocusing
mechanisms is investigated. The paper is concluded with a brief
discussion in Sec.~\ref{sec:Discussion}, and we present a number of
mathematical derivations in
Appendixes~\ref{app:Special_Lorentz}--\ref{sec:stark-shift-effects}.

\section{Physical modeling and equations of motion}
\label{sec:Eq_of_motion}
\begin{figure}[t]
  \centering
  \includegraphics{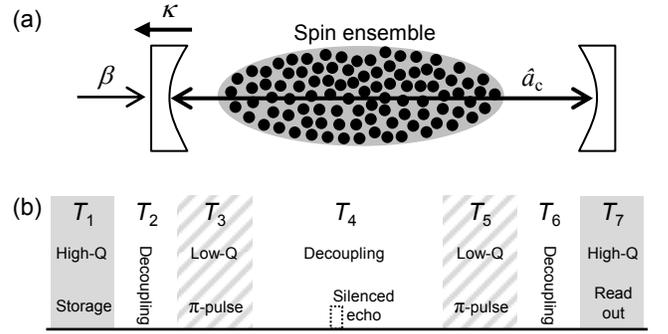}
  \caption{(a) The physical system under consideration: A spin
    ensemble is placed within a one-sided cavity, which may be driven
    externally by the field $\beta$ through a mirror with field-decay
    rate $\kappa$. (b) The quantum-memory protocol consists of storage
    and retrieval parts with a strongly and resonantly coupled
    spin-cavity system (shaded area), $\pi$-pulses for inverting the
    spin ensemble in a resonant but weakly coupled regime (hatched
    area), and parts (white) with effective spin-cavity
    decoupling. Due to the two $\pi$-pulses, a silenced spin echo
    occurs in the middle of the sequence. For later reference $T_i$
    denotes the duration of the $i$'th part of the sequence.}
  \label{fig:Setup}
\end{figure}

The physical system under consideration is shown schematically in
Fig.~\ref{fig:Setup}(a). The field $\ac$ in a one-sided cavity is
coupled to a spin ensemble, and the frequency $\omega_j$ of the $j$th
spin is assumed to be static but inhomogeneously broadened around a
central spin frequency $\omegas$ with a Lorentzian distribution:
\begin{equation}
\label{eq:f_Lorentz}
  f(\omega) = \frac{w/2\pi}{(\omega-\omegas)^2 + \frac{w^2}{4}},
\end{equation}
where $w$ is the full width at half maximum (FWHM). In the frame
rotating at $\omegas$ the free evolution of the spin ensemble and the
cavity field is governed by the Hamiltonian (taking $\hbar = 1$):
\begin{equation}
\label{eq:H_free}
  \H_0 = \Deltacs\adagc\ac + \sum_j\frac{\Delta_j}{2}\pauli_z^{(j)},  
\end{equation}
where $\Deltacs = \omegac - \omegas$ is the detuning of the cavity
resonance frequency $\omegac$, and $\Delta_j = \omega_j -
\omegas$. The coupling between the spin ensemble and the cavity field
is governed by the interaction Hamiltonian:
\begin{equation}
\label{eq:H_int}
  \Hint = \sum_j g_j(\pauli_+^{(j)}\ac+\pauli_-^{(j)}\adagc),
\end{equation}
where $g_j$ is the coupling strength of the $j$th spin and
$\pauli_+^{(j)}$, $\pauli_-^{(j)}$, and $\pauli_z^{(j)}$ are Pauli
operators. The cavity field can be driven by a coherent-state field
$\beta$ with the associated Hamiltonian:
\begin{equation}
  \label{eq:H_ext}
  \Hext = i\sqrt{2\kappa}(\beta \adagc - \beta^*\ac),
\end{equation}
where $\kappa$ is the field-decay rate of the cavity and $\beta$ is
normalized such that $|\beta|^2$ is the number of photons incident on
the cavity per second. Decay processes are handled in the Markov
approximation, e.g.~by the master equation:
$\frac{\partial\dens}{\partial t} = -i[\H,\dens] + \sum_n
\mathcal{L}[\c_n]\dens$, where $\H$ is the total Hamiltonian and the
Lindblad part is given by $\mathcal{L}[\c]\dens =
-\frac{1}{2}\cdag\c\dens - \frac{1}{2}\dens\cdag\c +
\cdag\dens\c$. The cavity leakage gives rise to a Lindblad term with
$\c_{\mathrm{c}} = \sqrt{2\kappa}\ac$, and for each individual spin a
collision-like dephasing with characteristic waiting time $\tau$ is
modeled by $\c_j = \frac{1}{\sqrt{2\tau}}\pauli_z^{(j)}$.

This manuscript applies the above formalism in two regimes: (i) In a
quantum-memory protocol the exchange of information between the cavity
field and the spin ensemble occurs in the linear regime with
$\mean{\pauli_z^{(j)}} \approx -1$, which can be described in the
Holstein-Primakoff approximation
\cite{Holstein.PhysRev.58.1098(1940)}. For the specific choice of
Lorentzian inhomogeneous broadening the dynamical evolution may often
be described exactly or approximately by analytical expressions (the
reason for this is discussed in appendix~\ref{app:Special_Lorentz}),
which enhances the physical understanding of the processes
involved. (ii) In order to employ spin-refocusing techniques in the
quantum-memory protocol, the spin ensemble must be subjected to
$\pi$-pulses which involves a non-linear regime of the dynamical
evolution. This can only be handled numerically; we shall employ a
method which divides the inhomogeneous spin ensemble into homogeneous
sub-ensembles, $\mathcal{M}_1, \mathcal{M}_2, \ldots, \mathcal{M}_M$:
\begin{equation}
  \label{eq:def_sub-ensembles}
  \S_x^{(m)}=\negthickspace\negthickspace\sum_{j\in \mathcal{M}_m}\pauli_x^{(j)},\quad
  \S_y^{(m)}=\negthickspace\negthickspace\sum_{j\in \mathcal{M}_m}\pauli_y^{(j)},\quad
  \S_z^{(m)}=\negthickspace\negthickspace\sum_{j\in \mathcal{M}_m}\pauli_z^{(j)},
\end{equation}
and which accounts for the quantum state through the first and second
moments of the physical variables (details can be found in
Ref.~\cite{Julsgaard.PhysRevLett.110.250503(2013)}). Comments on the
accuracy of the numerical procedure is presented in
Appendix~\ref{sec:Requirement_num_sim}.

\section{A spin-ensemble quantum memory: The basic idea}
\label{sec:Basic_idea}
The fundamental idea behind the spin-ensemble quantum memory relies on
the specific interaction Hamiltonian~(\ref{eq:H_int}), which can be
rewritten as: $\Hint = \gens(\ac\bdag + \adagc\b)$, where the
collective-spin-mode annihilation operator is given by:
\begin{equation}
\label{eq:def_b}
  \b = \sum_J \frac{g_j}{\gens}\pauli_-^{(j)},
\end{equation}
and $\gens = [\sum_j g_j^2]^{1/2}$ is the ensemble coupling constant
scaling as $\sqrt{N}$ times the individual-spin coupling strength. The
creation operator $\bdag$ is the hermitian conjugate of $\b$, and the
specific choice of $\gens$ ensures the commutation relation
$[\b,\bdag] = 1$ in the linear regime with $\mean{\pauli_z^{(j)}}
\approx -1$. On resonance ($\Deltacs = 0$), and in the absence of
inhomogeneities and decay mechanisms ($\Delta_j = \kappa = \tau^{-1} =
0$), the evolution is governed solely by $\Hint$ leading to the
following equations of motion in the Heisenberg picture:
\begin{equation}
\label{eq:Basic_spin_cavity_evolution}
  \begin{split}
    \ac(t) &= \cos(\gens t)\ac(0) -i\sin(\gens t)\b(0), \\
    \b(t) &= -i\sin(\gens t)\ac(0) + \cos(\gens t)\b(0).
  \end{split}
\end{equation}
At time $\Tswap = \frac{\pi}{2\gens}$ the cavity-field and
spin-ensemble quantum states are swapped. In a practical realization,
however, there are several challenges to address: (i) The
inhomogeneity in spin-resonance frequencies gives rise to a separate
phase evolution for the individual spins, $\pauli_-^{(j)}(t) =
\pauli_-^{(j)}(0) e^{-i\Delta_j t}$, and the spin-ensemble excitation
quickly disappears from the symmetric mode $\b$ coupled to the
cavity. (ii) This dephasing mechanism can be counter-acted by
spin-refocusing techniques, which however involves that the entire
spin ensemble must be transfered to the excited state with possible
instabilities as a concern
\cite{Julsgaard.PhysRevA.86.063810(2012)}. (iii) A mechanism must be
devised in order to effectively switch on and off the basic
interaction of Eq.~(\ref{eq:Basic_spin_cavity_evolution}). (iv) The
impact of decay processes (modeled by $\kappa$ and $\tau$) and their
interplay with the ensemble characteristics ($\gens$ and $w$) must be
understood in order to devise the best parameter regime. (v) The
spin-refocusing mechanisms may be driven by an external field $\beta$
through the cavity, and the feasibility of this process in terms of
energy or power must be accounted for.

The process in Eq.~(\ref{eq:Basic_spin_cavity_evolution}) has been
demonstrated experimentally in nitrogen-vacancy (NV) centers in
diamond coupled to a co-planar micro-wave cavity
\cite{Kubo.PhysRevLett.107.220501(2011)}. In that work it was not
attempted to employ spin-refocusing techniques, and inhomogeneous
broadening, indeed, presented the main limitation of the quantum
memory.

\subsection{A quantum memory protocol enabled by spin-refocusing
  techniques}
\label{sec:Protocol_general}
In this manuscript it is assumed that an initial state of the cavity
field is given, and our task is to transfer this state to the spin
ensemble and retrieve it again after a specified memory time,
$\Tmem$. To reach this goal we employ a quantum-memory protocol, the
most important constituents of which have been shown schematically in
Fig.~\ref{fig:Setup}(b). First, by a storage process the cavity field
must be transfered to the spin ensemble. During this process the
cavity leakage should be minimized, i.e.~the cavity-$Q$ parameter must
be high in order that the cavity-field decay rate is much slower than
the characteristic transfer rate. The same applies for the retrieval
process at the end of the protocol. Second, in order to employ
spin-refocusing techniques the spin ensemble must reside in an
inverted state (roughly) half of the memory time and near the ground
state for the remaining time, and two $\pi$-pulses are applied in
order to facilitate this. To prevent super-radiant processes during
these pulses and while the spins are excited, the cavity must be in a
low-$Q$ mode to ensure stability
\cite{Julsgaard.PhysRevA.86.063810(2012)}. Third, in between storage,
retrieval, and $\pi$-pulses the spin-cavity should evolve freely for
durations adjusted by the requirement that the spin-refocusing process
matches the desired memory time. In these periods the spin-cavity
system should be decoupled to prevent leakage of the stored quantum
state and to prevent generation of excess noise. We shall decouple the
spin-cavity system effectively by detuning the cavity frequency from
the spin-resonance frequency. Finally, between the main parts of the
protocol shown in Fig.~\ref{fig:Setup}(b) there are short periods of
time during which the cavity parameters, $\kappa$ and $\omegas$, are
adjusted. In this work they are treated as infinitely fast and are
thus disregarded. See
Ref.~\cite{Julsgaard.PhysRevLett.110.250503(2013)} for further details
of a practical memory protocol and of its experimental feasibility.

According to Eq.~(\ref{eq:Basic_spin_cavity_evolution}) an ideal
memory swaps the state of a cavity field into and back from the spin
ensemble oscillator according to: $\Xa^{\mathrm{out}} =
-\Xa^{\mathrm{in}}$ and $\Pa^{\mathrm{out}} = -\Pa^{\mathrm{in}}$,
where
\begin{equation}
  \Xa = \frac{\ac+\adagc}{\sqrt{2}},\qquad
  \Pa = \frac{-i(\ac-\adagc)}{\sqrt{2}},
\end{equation}
fulfill the commutation relation $[\Xa,\Pa] = i$. For this reason we
define the gain $\mathcal{G}$ in terms of the mean values of the
actual process as:
\begin{equation}
\label{eq:Def_gain_simple}
  X_{\mathrm{c}}^{\mathrm{out}} = -\mathcal{G}X_{\mathrm{c}}^{\mathrm{in}}, \qquad
  P_{\mathrm{c}}^{\mathrm{out}} = -\mathcal{G}P_{\mathrm{c}}^{\mathrm{in}}.
\end{equation}
We also note that the minimum uncertainty state of the cavity field
fulfills $\mean{\delta\Xa^2} = \mean{\delta\Pa^2} = \frac{1}{2}$, which
holds in particular for all coherent states. Hence, when applying
coherent input states to the quantum memory, the variance of the
output state quadratures $\sigma^2 \equiv
\mean{\delta\Xa^{\mathrm{out}\;2}} =
\mean{\delta\Pa^{\mathrm{out}\;2}} \ge \frac{1}{2}$ is a measure of
the added noise from the memory protocol. 

Due to the operator nature of
Eq.~(\ref{eq:Basic_spin_cavity_evolution}) the quantum memory should
work ideally for any quantum state, and the performance of the memory
protocol is fully characterized by its impact on coherent states due
to their (over)completeness. As will be quantitatively justified in
Sec.~\ref{sec:Asymmetric_in_out}, the protocol is well described in
practice by a linear input-output relation, and in this case the
transformation of the first and second moments of $\Xa$ and $\Pa$,
parametrized through $\mathcal{G}$ and $\sigma^2$, is sufficient for
characterizing the quantum memory performance \cite{FidelityPaper}. We
note that the protocol may give rise to phase rotations and
asymmetries in the above input-output relations, and in this case the
above definitions of $\mathcal{G}$ and $\sigma^2$ must be generalized,
which is the subject of Appendix~\ref{app:GainVarFidelity}.

\section{Swapping between the cavity field and spin components}
\label{sec:Spin-cavity-swap}
\begin{figure}[t]
  \centering
  \includegraphics{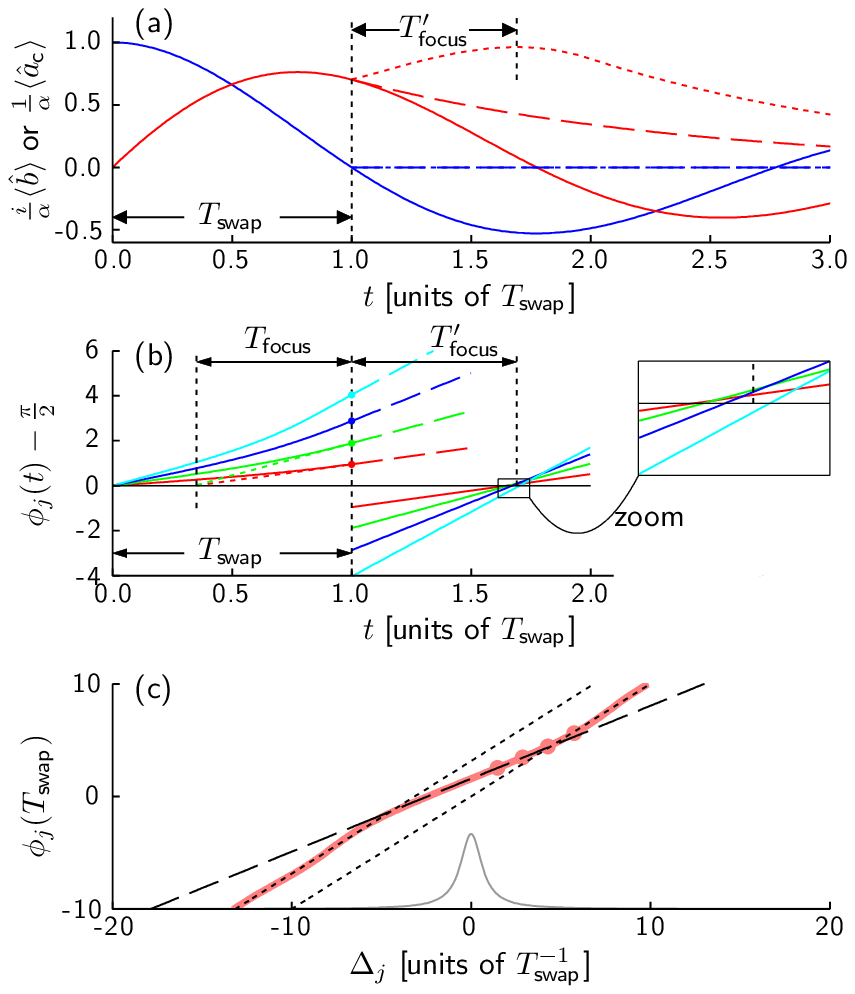}
  \caption{(a) The solid curves show the evolution of $\mean{\ac}$
    (blue) and $\mean{\b}$ (red) versus time for a coupled spin-cavity
    system with $\gens = 2.5\Gamma$. The dashed lines show the free
    evolution if $\gens$ is changed to zero at $t=\Tswap$, and the
    dotted lines show the same but with a perfect inversion process
    included at $t = \Tswap$ (the blue dashed and dotted lines
    coincide). (b) The solid curves show the phase $\phi_j$ of
    individual spin components for four different frequency classes
    with $\Delta_j$ taking the values (counting from the horizontal
    axis and away): $w$ (red), $2w$ (green), $3w$ (blue), and $4w$
    (cyan). The inversion process implies $\phi_j \rightarrow -\phi_j$
    at $t=\Tswap$ with the dashed lines representing the free
    evolution in the absence of the inversion process and the dotted
    lines (for the two lowest values of $\Delta_j$) the hypothetical
    free evolution of duration $\Tfocus$ back in time to the
    \emph{effective} point of origin. In panel (c) the thick solid
    line (red) shows the phase $\phi_j$ versus $\Delta_j$ at
    $t=\Tswap$---the four dots correspond to the four cases examined
    in (b). The dashed line represents a fit of $\Tfocus$ to the
    function $\phi_j = \frac{\pi}{2} + \Delta_j \Tfocus$ in the linear
    regime near $\Delta_j = 0$, and the dotted lines follow $\phi_j =
    \Delta_j\Tswap$ and $\phi_j = \Delta_j\Tswap + \pi$. The thin gray
    curve represents (on an arbitrary vertical scale) the
    inhomogeneous distribution of Eq.~(\ref{eq:f_Lorentz}) used in
    this example.}
  \label{fig:SwapIllustration}
\end{figure}
This section considers the impact of inhomogeneous broadening and
decay mechanisms on the otherwise idealized spin-cavity evolution of
Eq.~(\ref{eq:Basic_spin_cavity_evolution}). We start with a discussion
of mean values while second moments are covered in the end of the
section.

For the storage part of the quantum memory protocol, the relevant
initial state is the ground state for the spin ensemble,
$\mean{\pauli_z^{(j)}(0)} = -1$ and $\mean{\pauli_{\pm}^{(j)}(0)} =
0$, i.e.~$\mean{\b(0)} = 0$, while we take the cavity-field to be in a
coherent state of amplitude $\mean{\ac(0)} = \alpha$. With the cavity
coupled resonantly to the spin ensemble ($\Deltacs = 0$) the
subsequent evolution follows Eqs.~(\ref{eq:ac_during_swap})
and~(\ref{eq:b_during_swap}) as exemplified by the solid curves in
Fig.~\ref{fig:SwapIllustration}(a). In comparison to the idealized
behavior of Eq.~(\ref{eq:Basic_spin_cavity_evolution}) we now observe
an exponentially decaying envelope function
$\exp(-\frac{1}{2}[\kappa+\Gamma]t)$, where $\Gamma = w +
\frac{1}{2\tau}$, and also a slight slow-down of the oscillatory rate
$\gens \rightarrow \gens'$, with $\gens'$ being stated after
Eq.~(\ref{eq:b_during_swap}). The swapping time $\Tswap$ is defined by
the requirement $\mean{\ac(\Tswap)} = 0$, which is given analytically
by Eq.~(\ref{eq:Tswap_analytical}).

The properties of the spin state immediately after the resonant swap
process, $t=\Tswap$, is given not only by the specific spin-mode
$\b$---due to the spin-frequency inhomogeneity the stored information
is distributed among other spin modes already during the swapping
part, and the exact details of this distribution will eventually
affect the quantum-memory fidelity. The spin-state mean values at
$t=\Tswap$ can in principle be calculated by formal integration,
$\mean{\pauli_-^{(j)}(\Tswap)} =
-ig_j\int_0^{\Tswap}e^{-(\gammaperp+i\Delta_j)(\Tswap-t')}\mean{\ac(t')}dt'$
using Eq.~(\ref{eq:ac_during_swap}) where $\gammaperp = \tau^{-1}$. It
is more instructive, however, to examine the \emph{free} evolution of
spins for $t \ge \Tswap$, and to this end we artificially decouple the
spin-cavity system at $t=\Tswap$ by setting $g_j \rightarrow 0$. The
subsequent evolution of $\mean{\ac}$ and $\mean{\b}$ is exemplified in
Fig.~\ref{fig:SwapIllustration}(a) with dashed curves. While
$\mean{\ac}$ remains constant at zero, the spin mode $\mean{\b}$
decays due to the dephasing from inhomogeneous broadening. However, if
we at $t=\Tswap$ also impose an ideal inversion process (around the
$y$-axis), $\pauli_z^{(j)} \rightarrow -\pauli_z^{(j)}$ and
$\pauli_-^{(j)} = \frac{1}{2}(\pauli_x^{(j)} - i \pauli_y^{(j)})
\rightarrow \frac{1}{2}(-\pauli_x^{(j)} - i \pauli_y^{(j)})$, the mean
value of $\b$ is seen to increase and reach a maximum after a duration
$\Tfocus'$ due to rephasing of the spin ensemble (dotted curves in
Fig.~\ref{fig:SwapIllustration}(a)). Clearly, $\Tfocus'$ is smaller
than $\Tswap$, the physical reason being that the stored information
resides only part of the time in spin-degrees of freedom and
$\Tfocus'$ is an average measure on how much phase the individual
spins have accumulated during $\Tswap$. In order to illustrate the
evolution of individual spins, we plot in
Fig.~\ref{fig:SwapIllustration}(b) the phase $\phi_j$ for various
spin-frequency classes, where the phases are defined as
$\mean{\pauli_-^{(j)}} \equiv |\mean{\pauli_-^{(j)}}|
e^{-i\phi_j}$. The solid curves represent the scenario of decoupling
($g_j \rightarrow 0$) and ideal inversion immediately after the
swapping process at $t=\Tswap$, and the spin echo occurs when the
phases refocus at $\phi_j-\frac{\pi}{2} \approx 0$. However, as is
evident from the zoom-in panel, the different classes are refocused at
slightly different times, which eventually leads to a slight
degradation in quantum-memory performance. To elucidate this
phenomenon even further, the accumulated phase $\phi_j$ during the
swap is shown versus $\Delta_j$ in
Fig.~\ref{fig:SwapIllustration}(c). Clearly, for small values of
$|\Delta_j|$ there is a linear dependence and we define the slope to
be $\Tfocus$ (i.e.~the duration for refocusing in this linear regime)
being represented by the dashed line. This slope depends in a
non-trivial way on the inhomogeneous distribution, the coupling
constants, and the decay processes; however, a crude estimate $\Tfocus
\approx \frac{2}{3}\Tswap$ can be made, see
Eq.~(\ref{eq:pauliMinus_Swap_small_Deltaj}). For large values of
$|\Delta_j|$ another slope equal to $\Tswap$ occurs (dotted lines in
Fig.~\ref{fig:SwapIllustration}(c), see
Eq.~(\ref{eq:pauliMinus_Swap_large_Deltaj})), and the actual phase is
seen to change smoothly between these two regimes, which are
distinguished by $|\Delta_j|$ being either smaller or larger than the
characteristic rate $\Tswap^{-1}$ of the cavity-field variations.  The
above observations resemble the role of aberrations in geometric
optics, where e.g.~rays far from the center of lenses give rise to
imperfect imaging properties. In our case the magnitude of
imperfections is governed by the width of the linear regime compared
to the inhomogeneous frequency distribution (thin gray curve in
Fig.~\ref{fig:SwapIllustration}(c)) of spins storing the
information---we shall be more quantitative on this below.

\begin{figure}[t]
  \centering
  \includegraphics{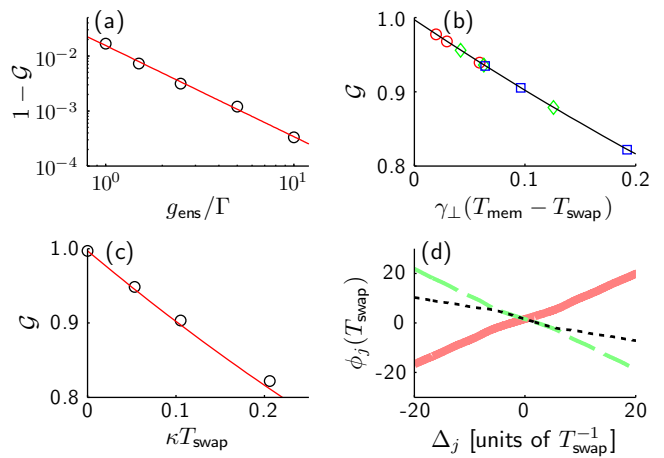}
  \caption{(a) The mean-value gain with $\kappa = \gammaperp = 0$ and
    varying $\gens$. The red line, for guiding the eye, has a slope of
    $\approx -1.7$ in the double-logarithmic plot. (b) The gain for
    varying values of $\gammaperp$ and $\Tmem$ with $\gens =
    2.5\Gamma$, the different symbols represent different values of
    $\Tmem$. The solid curve is $y = 0.997\exp(-x)$. (c) The gain
    versus $\kappa$ with $\gammaperp = 0$ and $\gens = 2.5\Gamma$. The
    solid line is $y = 0.997\exp(-x)$. (d) For the right-most data
    point in panel (c) the red solid curve shows the phase $\phi_j$
    versus $\Delta_j$ after the initial swap (in analogy to
    Fig.~\ref{fig:SwapIllustration}(c)). The green dashed curve is a
    time-reversed swap, i.e.~the desired target phase distribution
    before the retrieval part of the memory. The black dotted line is
    the actual phase distribution obtained by the refocusing process
    matching the desired slope for small $|\Delta_j|$.}
  \label{fig:SwapPerformance}
\end{figure}
In order to calculate the impact of the storage and retrieval part on
a quantum-memory protocol we employ the following idealized scenario:
(1) The storage and retrieval parts have durations (referring to
Fig.~\ref{fig:Setup}(b)) $T_1 = T_7 = \Tswap$ defined by the condition
$\mean{\ac(\Tswap)} = 0$. (2) For the decoupled parts, and during the
$\pi$-pulses, we set $g_j = 0$ such that the decoupling is ideal. (3)
The inversion pulses are perfect and infinitely fast. (4) The duration
of the three decoupling parts are adjusted such that $T_2 = T_6$,
$\sum_{i=1}^7T_i = \Tmem$, and $2\Tfocus + \sum_{i=2}^6 T_i = 2T_4$
(the latter ensures an even amount of time spent in the inverted and
non-inverted states between the effective focus points). The choice of
equal swapping times and focusing times for the storage and retrieval
parts is not obvious and will be commented on below. We note that for
$\gammaperp = 0$ the decoupling parts just serve as delay with no
degradation in memory performance.

Now, consider Fig.~\ref{fig:SwapPerformance}(a) which shows the gain
degradation versus $\gens/\Gamma$. Evidently, the performance improves
when this ratio increases; for larger $\gens$ the swap process occurs
faster, and in turn a larger fraction of the spin ensemble remains in
the linear region (i.e.~their phase is determined by the coupling to
the field and not by their own frequency during the swap). For
increasing $\gens/\Gamma$ the central linear regime in
Fig.~\ref{fig:SwapIllustration}(c) grows in comparison to the width of
the Lorentzian distribution.

Next, for a fixed ratio $\gens/\Gamma = 2.5$ leading to the gain
$\mathcal{G}_0 = 0.997$ in Fig.~\ref{fig:SwapPerformance}(a), we wish
to examine the impact of non-zero $\kappa$ and $\gammaperp$. From the
functional behavior of Eqs.~(\ref{eq:ac_during_swap})
and~(\ref{eq:b_during_swap}), i.e.~from the envelope function
$e^{-\frac{1}{2}(\kappa+\gammaperp)t}$ in the homogeneous case, we
make a naive guess that during the total storage and retrieval time,
$2\Tswap$, the gain degradation amounts to
$e^{-(\kappa+\gammaperp)\Tswap}$ while only $\gammaperp$ plays a role
in between according to $e^{-\gammaperp(\Tmem - 2\Tswap)}$. Hence, we
expect $\mathcal{G} = \mathcal{G}_0
e^{-\kappa\Tswap}e^{-\gammaperp(\Tmem-\Tswap)}$, which is examined by
the numerical results presented in
Fig.~\ref{fig:SwapPerformance}(b+c). In panel (b) the cavity leakage
is absent, $\kappa = 0$, while both $\gammaperp$ and $\Tmem$ are
varied. In panel (c) the spin decoherence is turned off, $\gammaperp =
0$, while $\kappa$ is varied. In both panels the solid curve
represents the above expectation and corresponds to the numerical
results to a good approximation. For this reason we wish to maintain
the above naive but simple expectation as a convenient rule of thumb.

The above considerations, however, do not represent the optimum
solution, which must in practice be calculated numerically. To
illuminate the underlying problems, consider first the swapping time,
$\Tswap$, which we took as equal for the storage and retrieval
parts. In the presence of decay mechanisms ($\kappa$ and $\gammaperp$
not both zero) there is actually an asymmetry in the storage and
retrieval part: The perfect retrieval consists of a cavity field
$\mean{\ac(\Tmem)} = -\mathcal{G}\alpha$ traced backward in time
until $\mean{\ac(\Tmem-\Tswap^{\mathrm{rev}})} = 0$, but the reversed
direction of time leads to an exponentially increasing behavior due to
$\kappa$ and $\gammaperp$, and hence the ``reverse swapping time'',
$\Tswap^{\mathrm{rev}}$ given by Eq.~(\ref{eq:Tswap_analytical}) with
sign changes on $\kappa$ and $\gammaperp$, is not exactly equal to the
forward $\Tswap$. Second, in the same manner the focusing time is also
changed, $\Tfocus \rightarrow \Tfocus^{\mathrm{rev}}$, and we may wish
to test the implications by the following protocol: (1) The storage
and retrieval times are set to $T_1 = \Tswap$ and $T_7 =
\Tswap^{\mathrm{rev}}$, (2) the decoupling periods are adjusted
according to $\Tfocus+\Tfocus^{\mathrm{rev}} + \sum_{i=2}^6 T_i =
2T_4$ instead of the choice made above. The phases $\phi_j$ of
individual spin components then follow the red solid line in
Fig.~\ref{fig:SwapPerformance}(d) after the initial storage, and the
time-reversed evolution of the perfect retrieval amounts to the green
dashed line (the example here corresponds to the parameters of the
right-most data point in Fig.~\ref{fig:SwapPerformance}(c),
i.e.~$\gammaperp = 0$ and $\kappa \ne 0$). Now, the mathematical
implications of the refocusing mechanism is to change the slope of the
red solid line reaching the black dotted curve. The particular usage
of $\Tfocus$ and $\Tfocus^{\mathrm{rev}}$ ensures a perfect match of
slopes in the central part with small $|\Delta_j|$; however, the
entire frequency range cannot be matched, and it is intuitively clear
that slight changes in the slope of either the dashed or dotted curves
in Fig.~\ref{fig:SwapPerformance}(d) might improve the
performance. The gain resulting from the scenario of panel (d) is in
fact 2 \% lower than the value found in panel (c). Although the
choices behind the result of Fig.~\ref{fig:SwapPerformance}(a,b,c) are
not reached by a true optimization, it is simple, pragmatic, and works
very well. For this reason we maintain this choice for the remaining
of the manuscript.

Turning to the second moments, we expect both the cavity field and the
spin components to remain in their minimum uncertainty state. Excess
noise is generated due to our inability to predict whether excitations
reside in the cavity or in the spin ensemble. For a non-inverted
ensemble there is no energy available to facilitate any unknown
distribution of excitations (the energy represented by the input
quantum field is weak and also insufficient in this respect), and
hence we expect no excess noise generated in the storage and retrieval
processes. For an inverted state we know that excess noise may be
generated \cite{Julsgaard.PhysRevA.86.063810(2012)}; however, for the
specific case discussed here with $g_j = 0$ during periods of
inversion, there is no exchange of energy between the spin ensemble
and the cavity field, and in turn our knowledge of the excitations is
not degraded. In total, no excess noise is expected in the idealized
protocol, and in particular, the storage and retrieval processes do
not generate noise. We have confirmed this by numerical simulations.

\section{Spin-cavity decoupling by frequency detuning}
\label{sec:spin-cavity-decoupling}
The decoupling of the spin ensemble from the cavity field during parts
of the quantum memory protocol, see Fig.~\ref{fig:Setup}(b), is
necessary for the following reasons: (i) The primary echo occurring
half-way through the protocol should be silenced
\cite{Damon.NewJPhys.13.093031(2011)}, i.e.~the energy represented by
the spin excitation must not leak to the cavity. (ii) During periods
of spin-ensemble inversion the spin-cavity coupling generates excess
noise \cite{Julsgaard.PhysRevA.86.063810(2012)}, and a proper
decoupling prevents this excess noise from interfering with the
particular spin-mode holding the stored quantum state. (iii) The
standard Hahn-spin-echo scheme employed here is based on the idea that
the individual \emph{free} spin evolution, $\pauli_-(t) = \pauli_-(0)
e^{-i\Delta_j t}$, can be reversed by appropriate spin-inversion
processes such that the total phase accumulated by each spin is
independent on $\Delta_j$.

We note that there are physical implementations, in which a direct,
on-demand decoupling, $g_j \rightarrow 0$, is possible, e.g.~in atomic
lambda systems with the cavity field coupled to the spin ensemble by a
Raman process \cite{Black.PhysRevLett.95.133601(2005)}. However, in
this manuscript the decoupling mechanism is based on detuning the
cavity frequency from the spin-resonance frequency, and the present
section describes quantitatively the implications of this detuning,
$\Deltacs$, being finite. Physically, the spin-ensemble oscillator
remains coupled to the cavity-field oscillator. Thus, in connection to
point (ii) above, a small probability remains for exchanging energy
between the two oscillators, and hence an increased noise in the spin
and cavity variables. In connection to point (iii), the spin-ensemble
oscillator may induce a small cavity field, which in turn presents a
slight back action on the otherwise free spin evolution. This is
equivalent to the ac Stark effect on electrical dipoles, and the
effect is discussed mathematically in
Appendix~\ref{sec:stark-shift-effects}.

In order to isolate and investigate the impact of the \emph{effective}
spin-cavity decoupling on the quantum memory protocol, we numerically
simulate the following scenario: The initial coherent state is
prepared directly in the spin-ensemble oscillator, $\mean{\ac(0)} =
0$ and $\mean{\b(0)} = b_0$, i.e.~the swapping procedures (parts 1 and
7 in Fig.~\ref{fig:Setup}(b)) discussed in
Sec.~\ref{sec:Spin-cavity-swap} are absent. In addition, the
spin-inversion processes (parts 3 and 5 in Fig.~\ref{fig:Setup}(b))
are ideal and infinitely fast. Hence, we essentially model a pure
spin-refocusing process with $T_2 = T_6 \equiv T$, $T_4 = 2T$, and
$\Tmem = 4T$ but with the cavity playing a ``spectator role''. During
parts 2 and 6 the detuning is $\Deltacs$, and during part 4 it is
either of $\Deltacs' = \pm \Deltacs$. For simplicity we keep $\kappa$
constant in the entire period.

\begin{figure}[t]
  \centering
  \includegraphics{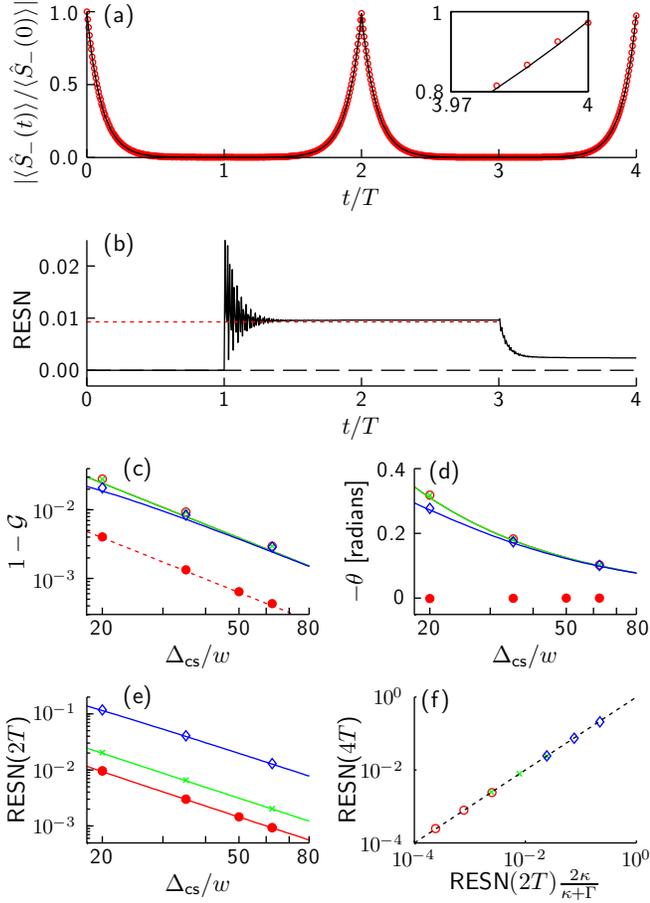}
  \caption{All panels: $\gammaperp = 0$ and $\gens = 2.5\Gamma$. (a)
    The transverse-spin-component mean value versus time for the
    spin-refocusing protocol of Sec.~\ref{sec:spin-cavity-decoupling}
    with $\kappa/w = 0.075$ and $\Deltacs/w = 20$. Red circles:
    Numerical simulation. Black solid line: Adiabatic model of
    Eqs.~(\ref{eq:Sminus_elim_cav_v1},
    \ref{eq:Sminus_v2_cav_elim_all_orders}-\ref{eq:Sminus_cav_elim_after_4T}).
    (b) For the same parameters as panel (a) the solid line shows the
    relative excess spin noise (RESN) versus time. The red dotted line
    denotes the steady-state value of
    Eq.~(\ref{eq:SpinRef_ConstDet_gain_RESN_mid}). (c) $1-\mathcal{G}$
    versus spin-cavity detuning $\Deltacs$ for $\kappa/w$ taking the
    values of 0.075 (red circles), 0.75 (green crosses), and 7.5 (blue
    diamonds). Open symbols correspond to $\Deltacs' = \Deltacs$ while
    the closed symbols (circles only) are obtained with $\Deltacs' =
    -\Deltacs$. Solid lines show the prediction of
    Eq.~(\ref{eq:SpinRef_ConstDet_gain}) while the red dotted line
    corresponds to $1-\mathcal{G} = \gens^2/\Deltacs^2$. (d) With the
    same symbols as panel (c) the memory phase shift versus
    $\Deltacs$. Solid lines correspond to
    Eq.~(\ref{eq:SpinRef_ConstDet_theta}). (e) The RESN at the mid
    point, $t=2T$, versus $\Deltacs$ with symbols of panel (c). The
    open and closed circles are superimposed. Solid lines are
    theoretical according to
    Eq.~(\ref{eq:SpinRef_ConstDet_gain_RESN_mid}). (f) The relation
    between the RESN at the mid- and end-points of the protocol. All
    data points were obtained for $\tilde{C} < 0.06$ and shown with
    the symbols of panel (c).}
  \label{fig:DecouplingProperties}
\end{figure}
An example of numerical simulations for the above protocol with
$\Deltacs' = \Deltacs$ is shown in
Fig.~\ref{fig:DecouplingProperties}(a) for mean values of the spin and
in Fig.~\ref{fig:DecouplingProperties}(b) for the relative excess spin
noise, RESN = $\frac{1}{2N}\mean{\delta\S_x^2 + \delta\S_y^2}-1$ (we
remind that the coherent-state variance is $\mean{\delta\S_x^2} =
\mean{\delta\S_y^2} = N$). With the protocol gain $\mathcal{G}$ and
phase shift $\theta$ defined by $\mean{\S_-(\Tmem)} =
\mean{\S_-(0)}\cdot \mathcal{G}e^{-i\theta}$, we see by the zoom-in of
the inset in panel (a) that the gain is slightly below unity at
$t=\Tmem$. In this example the spin decoherence rate is absent,
$\gammaperp = 0$, and the fact that $\mathcal{G} < 1$ shows that the
spin-cavity decoupling is not completely ideal. It is possible to
predict approximately the behavior of the mean values since the large
spin-cavity detuning, $\Deltacs \gg \gens$, allows adiabatic
elimination of the cavity field from the dynamical equations, see
appendix~\ref{sec:stark-shift-effects} for details. The solid lines of
panel (a) are examples of such predictions, and the gain and phase
shift are expected to be:
\begin{align}
\label{eq:SpinRef_ConstDet_gain}
  \mathcal{G} &\approx e^{-\gammaperp\Tmem}\left(1 - \left[\frac{\gens^2(\Deltacs+
   \Deltacs')/w}{\kappa^2+\Deltacs^2}\right]^2\right), \\
\label{eq:SpinRef_ConstDet_theta}
  \theta &\approx -2\arctan\left[\frac{\gens^2(\Deltacs+
   \Deltacs')/w}{\kappa^2+\Deltacs^2}\right].
\end{align}
In Fig.~\ref{fig:DecouplingProperties}(c) the numerically determined
gain (open symbols with $\Deltacs' = \Deltacs$) is compared to the
above expression (solid lines), and the agreement is seen to be quite
good (the symbols being slightly higher than the solid lines). We note
that Eq.~(\ref{eq:SpinRef_ConstDet_gain}) predicts unity gain when
$\Deltacs' = -\Deltacs$, which is not confirmed by the numerical
calculations (closed symbols in
Fig.~\ref{fig:DecouplingProperties}(c)). The difference between the
open red symbols and the corresponding solid line matches quite well
the residual gain imperfection represented by the closed symbols,
which are seen to scale as $\gens^2/\Deltacs^2$ (dotted red line). We
note that in steady state $\mean{\ac} =
\frac{-i\gens\mean{\b}}{\kappa+i\Deltacs}$ such that the relative
energy leakage to the cavity at $t=\Tmem$, being of the order
$\gens^2/\Deltacs^2$, is missing from the spin degree of freedom. This
effect was not covered by the adiabatic theory of
appendix~\ref{sec:stark-shift-effects}. We also examine the memory
phases shift $\theta$ in Fig.~\ref{fig:DecouplingProperties}(d), which
is seen to agree very well with the prediction of
Eq.~(\ref{eq:SpinRef_ConstDet_theta}).

Now, turning to the variance of the spin components, all individual
spins are in the same state ($\mean{\pauli_z^{(j)}} = 1$,
$\mean{\pauli_-^{(j)}} = 0$) after the first ideal inversion pulse at
$t=T$ if we consider the initial quantum state as the vacuum state,
$b_0 = 0$ (a weak non-zero value does not change this picture). The
subsequent evolution can then in principle be computed analytically by
an off-resonant version of the calculations in
appendix~\ref{sec:Requirement_num_sim}, which in steady state leads to
the following expression for the RESN:
\begin{equation}
\label{eq:SpinRef_ConstDet_gain_RESN_mid}
  \mathrm{RESN}(2T) = \frac{2\kappa\tilde{C}}{(\kappa+\Gamma)(1-\tilde{C})},
\end{equation}
where $\tilde{C} = C[1+\frac{{\Deltacs'}^2}{(\kappa+\Gamma)^2}]^{-1}$
generalizes the cooperativity parameter $C =
\frac{\gens^2}{\kappa\Gamma}$ derived for a resonant spin-cavity
interaction \cite{Julsgaard.PhysRevA.86.063810(2012)}. The above
expression is seen to agree very well with the numerically calculated
mid-point ($t=2T$) excess noise as shown in
Fig.~\ref{fig:DecouplingProperties}(e). After the second inversion
process at $t=3T$ we cannot expect the new steady-state value of spin
variances to be calculated from the simplified homogeneous case as in
appendix~\ref{sec:Requirement_num_sim} since at $t=3T$ the state of
individual spins is correlated to $\Delta_j$. However, due to the
refocusing effect around $t=3T$ the new steady-state spin variance
during $3T \le t \le 4T$ must be affected by the spin variance
recorded into the spin memory during $2T \le t \le 3T$. The RESN at
$t=4T$ depends in a complicated way on the RESN at $t=2T$, but in the
limit of efficient spin-cavity decoupling ($\tilde{C} \ll 1$) the
numerical simulations show that to a good approximation the two spin
variances fulfill: $\mathrm{RESN}(4T) =
\mathrm{RESN}(2T)\frac{2\kappa}{\kappa+\Gamma}$, see
Fig.~\ref{fig:DecouplingProperties}(f).

To conclude this section, the effective spin-cavity decoupling by
detuning leaves the spin ensemble to evolve almost freely provided
that the cavity-induced phase shift from various periods of finite
macroscopic spin polarization is balanced (e.g.~$\Deltacs > 0$ around
the peaks at $t=0$ and $t=4T$, and $\Deltacs < 0$ around the peak at
$t=2T$ in Fig.~\ref{fig:DecouplingProperties}(a) as discussed
above). In this spin-refocusing protocol, a small residual effect on
the gain scales as $1-\mathcal{G} \approx \frac{\gens^2}{\Deltacs^2}$
and a small excess variance is present, originating from the inverted
spin ensemble and scaling as $\mathrm{RESN}(\Tmem) = 2\sigma^2-1
\approx \frac{4\kappa\gens^2}{\Gamma\Deltacs^2}$ when $\tilde{C} \ll
1$ and $\Deltacs \gg \kappa+\Gamma$.

\section{Externally applied inversion pulses}
\label{sec:extern-appl-inversion}
\begin{figure}[t]
  \centering
  \includegraphics{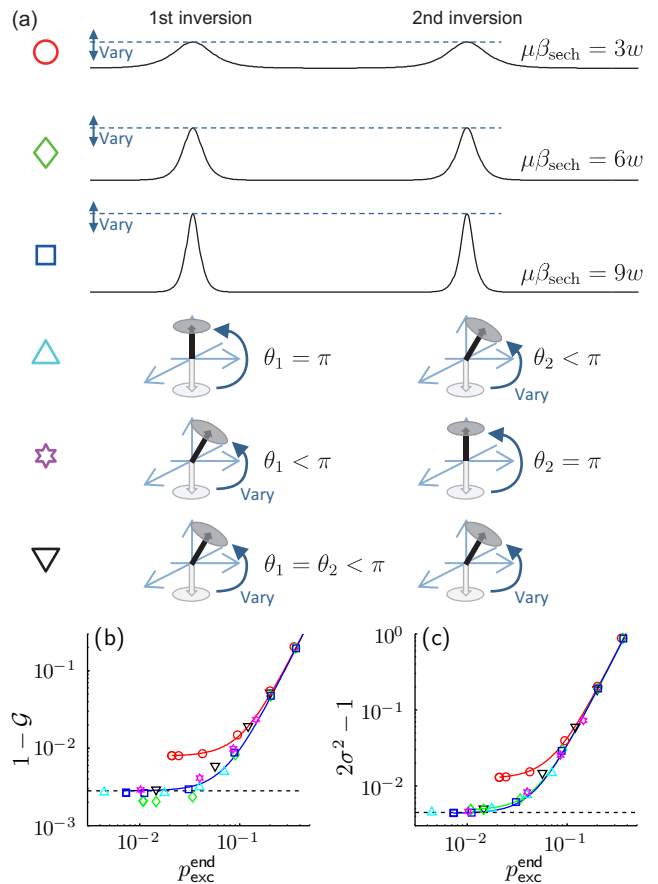}
  \caption{(a) Illustration of the inversion schemes investigated, the
    varied parameters, and the symbols used for plotting. The first
    three cases employ an externally driven intra-cavity field with
    secant hyperbolic shape. In the last three cases the entire spin
    ensemble is merely rotated by hand. (b) The average gain
    $\mathcal{G}$ versus the obtained excitation probability
    $p_{\mathrm{exc}}^{\mathrm{end}}$ at the end of the protocol. (c)
    The relative excess noise variance $\mathrm{REN} = 2\sigma^2-1$
    versus $p_{\mathrm{exc}}^{\mathrm{end}}$. In both panels (b) and
    (c) the solid lines are guides to the eye and the dashed
    horizontal lines correspond to ideal $\pi$-rotation for both
    inversion pulses.}
  \label{fig:Influence_of_inversion}
\end{figure}
We now turn to the investigation of how the inversion pulses affect
the performance of the memory protocol. Referring to
Fig.~\ref{fig:Setup}(b), we maintain for the storage and read-out a
perfect, non-leaking cavity (i.e.~infinite $Q$-parameter and $\kappa =
0$ during pulses 1 and 7). The decoupling pulses 2, 4, and 6 are also
kept in constant conditions with a fixed intermediate $Q$ ($\kappa =
0.75w$) and $\Deltacs = 50w$ for pulses 2 and 6 and $\Deltacs = -50w$
for pulse 4, such that the total cavity-induced phase shift is
minimized according to Sec.~\ref{sec:spin-cavity-decoupling}. Then,
only the nature of inversion pulses 3 and 5 are varied and the effect
on the memory performance is extracted. We shall employ two strategies
for these inversion pulses: (i) hyperbolic secant pulses
\cite{Silver.PhysRevA.31.R2753(1985)} for the cavity field, and (ii)
rotations ``by hand'' of the entire spin ensemble. The different
strategies have been shown schematically in
Fig.~\ref{fig:Influence_of_inversion}(a). As a starting point we
neglect spin dephasing ($\tau = \infty$).

The hyperbolic secant pulses are widely used and perform well for
inhomogeneous distributions of light-matter couplings
\cite{Garwood.JMagRes.153.155(2001)}. In this case the cavity field
varies as: $a_{\mathrm{c}}(t) =
a_{\mathrm{c}}^{\mathrm{max}}\mathrm{sech}(\beta_{\mathrm{sech}}
(t-t_{\mathrm{sech}}))^{1+i\mu}$, where
$a_{\mathrm{c}}^{\mathrm{max}}$ is the maximum amplitude of the cavity
field, $\beta_{\mathrm{sech}}^{-1}$ is the characteristic duration of
the pulse, $t_{\mathrm{sech}}$ is the center time of the pulse, and
$\mu$ determines the shape of the inversion frequency profile (the
higher $\mu$, the closer the profile resembles a top-hat distribution
of width $\mu\beta_{\mathrm{sech}}$). The maximum cavity field gives
rise to a maximum Rabi frequency $\chi^{\mathrm{max}} =
2ga_{\mathrm{c}}^{\mathrm{max}}$ (with $g$-being the coupling
parameter from the interaction Hamiltonian~(\ref{eq:H_int}) assumed to
be equal for all spins), and when $\chi^{\mathrm{max}} \ge
\mu\beta_{\mathrm{sech}}$ the inversion is known to work well. We will
vary the Rabi frequency across this threshold value and hence deduce
the effect of insufficient driving. In addition, various inversion
bandwidths $\mu\beta_{\mathrm{sech}}$ will be examined. We note that
the externally applied field $\beta$ must be tailored to account for
cavity filtering and for the reaction field of the spin dipoles
\cite{Julsgaard.PhysRevLett.110.250503(2013)}. The external driving is
truncated to a finite duration of $\approx 16/\beta_{\mathrm{sech}}$,
during which the cavity is tuned to low-$Q$ mode ($\kappa = 7.5w$).
The inversion bandwidths examined are $3w$, $6w$, and $9w$, for which
the resulting average gain $\mathcal{G}$ and relative excess variance
$2\sigma^2-1$ have been plotted in
Fig.~\ref{fig:Influence_of_inversion}(b,c). The varying strength of
the external driving $\beta$ leads to varying probabilities
$p_{\mathrm{exc}}^{\mathrm{end}}$ for a spin to be excited at the end
of the protocol, which is conveniently used as horizontal axis in
these figures. Evidently, for insufficient driving (leading to large
excitation probabilities $p_{\mathrm{exc}}^{\mathrm{end}}$) the
protocol performs poorly both in terms of gain and
variance. Conversely, when the external driving is sufficient, both
$\mathcal{G}$ and $2\sigma^2$ become quite close to their ideal values
of unity---the asymptotic values symbolized by the horizontal dashed
lines will be discussed shortly. For the case of
$\mu\beta_{\mathrm{sech}} = 3w$ (red circles in
Fig.~\ref{fig:Influence_of_inversion}) this asymptotic value is
significantly higher than in the remaining examples, which we
attribute to an insufficient frequency-bandwidth of the inversion
pulse (i.e.~a non-negligible fraction of spins are poorly inverted
despite a large driving strength).

Now, to examine whether the above results are general or unique to the
hyperbolic secant pulses, we employ a series of inversion pulses where
the entire spin ensemble is simply rotated abruptly by an angle equal
or close to the ideal value of $\pi$. The cavity is maintained in
low-$Q$ mode for a duration $\approx 12/w$ around the abrupt spin
rotation. We examine the cases where the first, second, and both
inversion pulses are non-perfect. In all three cases, a non-ideal
rotation angle $\ne \pi$ leads to a finite excitation probability
$p_{\mathrm{exc}}^{\mathrm{end}}$ at the end of the protocol. The
results have been added to Fig.~\ref{fig:Influence_of_inversion}, and
it is indeed seen that all the simulation results fall approximately
on the same curve. In other words, the final excitation probability
$p_{\mathrm{exc}}^{\mathrm{end}}$ defines the performance to a large
extent. The horizontal dashed lines correspond to the case where both
rotation angles have been set to $\pi$, leading essentially to
$p_{\mathrm{exc}}^{\mathrm{end}} = 0$. The reason that $\mathcal{G} <
1$ and $2\sigma^2 > 1$ despite the ideal $\pi$-pulse inversions was
discussed in Sec.~\ref{sec:spin-cavity-decoupling}.

\begin{figure}[t]
  \centering
  \includegraphics{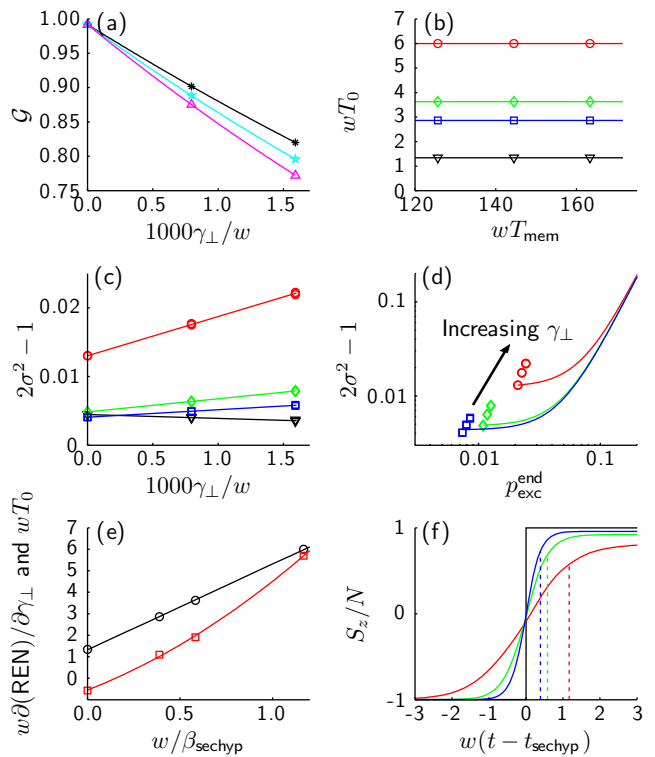}
  \caption{(a) Gain $\mathcal{G}$ versus $\gammaperp$ for
    $\mu\beta_{\mathrm{sech}} = 3w$ and various values of $w\Tmem =
    126$ (black asterisk), $146$ (cyan pentagram), and $163$ (magenta
    triangles). Solid lines represent fits to the function
    $\mathcal{G} = \mathcal{G}_0 e^{-\gammaperp(\Tmem - T_0)}$ with
    free parameters $\mathcal{G}_0$ and $T_0$. (b) The fitted $T_0$ is
    independent of $\Tmem$ for various values of
    $\mu\beta_{\mathrm{sech}}/w = 3$ (red circles), $6$ (green
    diamonds), $9$ (blue squares), and $\infty$ (black triangles). (c)
    The relative excess noise $\mathrm{REN} = 2\sigma^2-1$ versus
    $\gammaperp$ for various values of $\mu\beta_{\mathrm{sech}}$
    using the same symbols as in panel (b). (d) The effect of
    $\gammaperp$ on the REN (symbols) in comparison to the guides to
    the eye (solid lines) from
    Fig.~\ref{fig:Influence_of_inversion}(c). (e) Black circles: The
    fitted values of $T_0$ versus $\beta_{\mathrm{sech}}^{-1}$
    following the linear fit: $T_0 = 4.0/\beta_{\mathrm{sech}} +
    0.92\Tswap$. Red squares: The slopes of the lines from panel (c)
    showing the susceptibility of the REN to $\gammaperp$. (f)
    Behavior of the collective spin variable $S_z$ versus time during
    the inversion process for the values of $\beta_{\mathrm{sech}}$
    used in panels (b-e); the larger the $\beta_{\mathrm{sech}}$ the
    steeper the curve and the faster the inversion process.}
  \label{fig:Influence_of_gammeperp}
\end{figure}

\subsection{The effect of dephasing during inversion pulses}
\label{sec:Dephase_during_invert}
Our next step is to chose a finite dephasing time ($\tau < \infty$)
for the spin coherence. According to the discussions in
Sec.~\ref{sec:Spin-cavity-swap} we would naively expect the gain to
scale roughly as $\mathcal{G} \propto e^{-\gammaperp(\Tmem-\Tswap)}$,
but in the following we shall see that the inversion pulses impose
some corrections to this picture. The impact on the noise variance
will also be examined. As a starting point, the inversion strategy of
hyperbolic secant pulses (the first three strategies in
Fig.~\ref{fig:Influence_of_inversion}(a)) are compared for various
values of $\gammaperp$ and $\Tmem$ using a sufficient external driving
($\chi^{\mathrm{max}} = \mu\beta_{\mathrm{sech}}$). The performance of
the gain parameter $\mathcal{G}$ is shown in
Fig.~\ref{fig:Influence_of_gammeperp}(a), and the simulation results
can be fitted to the function $\mathcal{G} = \mathcal{G}_0
e^{-\gammaperp(\Tmem - T_0)}$; however, with a fitting parameter
$T_0$, which does not match the naive guess of $\Tswap$ but instead
varies with $\beta_{\mathrm{sech}}$ as illustrated in
Fig.~\ref{fig:Influence_of_gammeperp}(b): The smaller the
$\beta_{\mathrm{sech}}$, the longer the inversion process, and in turn
the longer the fitted $T_0$. The black triangles in
Fig.~\ref{fig:Influence_of_gammeperp}(b) corresponds to an infinitely
fast hyperbolic secant pulse and is calculated using two ideal
$\pi$-rotations as outlined by the sixth strategy in
Fig.~\ref{fig:Influence_of_inversion}(a). This relation between $T_0$
and $\beta_{\mathrm{sech}}$ is also shown with black circles in
Fig.~\ref{fig:Influence_of_gammeperp}(e), and we confirm the naive
guess $T_0 \approx \Tswap$ for very fast inversion pulses
only. Physically, the stored information resides partly as population
degrees of freedom during the inversion process, which in turn has a
shielding effect from the dephasing processes: The longer the duration
of the inversion pulses, the shorter becomes the effective time
$\Tmem-T_0$ of dephasing---the durations used in the simulations have
been illustrated in Fig.~\ref{fig:Influence_of_gammeperp}(f).

From the above discussion one may think that longer inversion pulses
may be slightly advantageous since the dephasing is partly turned
off. However, this effect comes at a price when considering the noise
properties of the memory protocol. While the simulations show that the
noise variance does not depend on $\Tmem$ (additional waiting time
does not contribute additional noise), it does depend on both
$\gammaperp$ and $\beta_{\mathrm{sech}}$ as shown in
Fig.~\ref{fig:Influence_of_gammeperp}(c). In general, dephasing
processes may counter-act noise generation [it dampens both mean
values and second moments in the equations of motion], which is indeed
seen by the black triangles in
Fig.~\ref{fig:Influence_of_gammeperp}(c) for infinitely fast inversion
pulses. However, turning to hyperbolic secant pulses there is an
additional noise generation, which grows with increasing
$\gammaperp$. This noise generation is more pronounced when the
characteristic duration $\beta_{\mathrm{sech}}^{-1}$ of the inversion
pulses is longer. In other words, while the damping of the gain
parameter $\mathcal{G}$ is reduced during the inversion pulse, there
is an accompanying noise generation at the same time. We note that
increasing $\gammaperp$ will decrease the quality of the inversion
process in terms of population---the hyperbolic secant pulses simply
perform worse in presence of dephasing. The effect of imperfect
inversion was investigated in
Fig.~\ref{fig:Influence_of_inversion}(c), showing approximately a
monotonous connection between $p_{\mathrm{exc}}^{\mathrm{end}}$ and
$2\sigma^2-1$. However, this effect cannot alone explain the increased
noise as is evident from Fig.~\ref{fig:Influence_of_gammeperp}(d);
although $p_{\mathrm{exc}}^{\mathrm{end}}$ increases slightly when
$\gammaperp$ grows, the increase in the relative excess noise variance
$2\sigma^2-1$ (symbols) is markedly larger than the population effect
corresponding to the guides-to-the-eye from
Fig.~\ref{fig:Influence_of_inversion}(c), which are reproduced in
Fig.~\ref{fig:Influence_of_gammeperp}(d). Hence, the increase in noise
caused by dephasing must arises during the inversion process. We
observe from Fig.~\ref{fig:Influence_of_gammeperp}(e), the relative
gain reduction [equal to
$\frac{1}{\mathcal{G}}\frac{\partial\mathcal{G}}{\partial(\gammaperp/w)}
= wT_0$ shown by black circles] is comparable to the relative noise
increase [equal to
$\frac{1}{2\sigma^2}\frac{\partial(2\sigma^2)}{\partial(\gammaperp/w)}
\approx w\frac{\partial(\mathrm{REN})}{\partial\gammaperp}$ shown by
red squares].
\begin{figure}[t]
  \centering
  \includegraphics{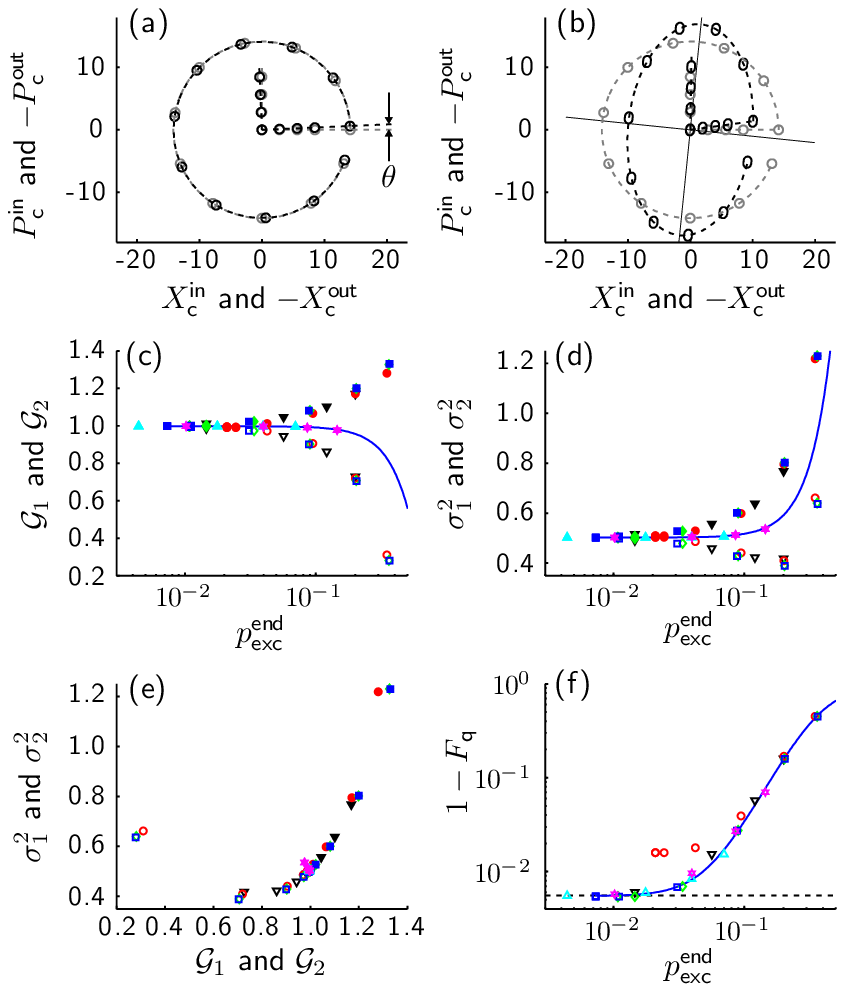}
  \caption{(a) Illustration of various input states (gray circles) and
    output states (black circles) for a secant hyperbolic excitation
    with $\chi^{\mathrm{max}} = \mu\beta_{\mathrm{sech}} = 9w$. The
    center of circles represent mean values of the cavity field
    quadratures $\Xa$ and $\Pa$, and the radius of the circles
    represent the square root of the variances $\mean{\delta\Xa^2}$
    and $\mean{\delta\Pa^2}$. (b) The same as panel (a) but with
    $\chi^{\mathrm{max}}$ reduced to
    $0.6\mu\beta_{\mathrm{sech}}$. Both gain and variances become
    asymmetric. (c) With the symbols defined in
    Fig.~\ref{fig:Influence_of_inversion} the gain along the minor
    axis (open symbols) and major axis (closed symbols) are shown
    versus the final excitation probability
    $p_{\mathrm{exc}}^{\mathrm{end}}$. (d) Variances along the minor
    axis (open symbols) and major axis (closed symbols) versus
    $p_{\mathrm{exc}}^{\mathrm{end}}$. (e) The relationship between
    gain and variances; closed symbols: $\sigma_1^2$ versus
    $\mathcal{G}_1$; open symbols: $\sigma_2^2$ versus
    $\mathcal{G}_2$. (f) Qubit infidelity versus
    $p_{\mathrm{exc}}^{\mathrm{end}}$. The blue solid lines in panels
    (c,d,f) correspond to the guide-to-the-eye already shown in
    Fig.~\ref{fig:Influence_of_inversion}.}
  \label{fig:AsymmetricGainAndVar}
\end{figure}

\subsection{Asymmetric input-output relations}
\label{sec:Asymmetric_in_out}
As our final discussion of inversion pulses, we consider the actual
input-output map of the memory protocol. This map was introduced in
simple terms in Eq.~(\ref{eq:Def_gain_simple}), but it turns out that
a generalization is required when inversion pulses are of
insufficient strength. Consider first
Fig.~\ref{fig:AsymmetricGainAndVar}(a) which shows the mean value and
standard deviation of $\Xa$ and $\Pa$ for both input and output states
in a simulation series using hyperbolic secant pulses of sufficient
strength and bandwidth. Each gray circle denote by its center the mean
value of $\Xa^{\mathrm{in}}$ and $\Pa^{\mathrm{in}}$, and its radius
corresponds to the standard deviations
$\mean{\delta\Xa^{\mathrm{in}\,2}}^{1/2} =
\mean{\delta\Pa^{\mathrm{in}\,2}}^{1/2}$. The black circles denote the
same for the output state, and apart from a tiny phase rotation
$\theta$ (caused by the cavity-induced phase shifts being not
completely compensated) the gray and black circles are essentially
equal, i.e. $\mathcal{G} \approx 1$ and no extra noise is added. In
contrast, the example given in Fig.~\ref{fig:AsymmetricGainAndVar}(b)
shows a more complicated input-output map, which mathematically
follows the parametrization of Eq.~(\ref{eq:Def_gain_general}). In
addition to an overall phase shift induced by the spin-cavity
coupling, this generalized map allows for describing the situation in
which the gain parameter is not symmetric in $\Xa,\Pa$-space but
instead depends on the phase of the input state. We note that the
inversion pulses are encoded with a specific phase, which breaks the
symmetry. The input-output map is decomposed into two main axes; a
major axis of gain $\mathcal{G}_1$ and variance $\sigma_1^2$ and a
minor axis of gain $\mathcal{G}_2$ and variance $\sigma_2^2$. The
asymmetric transformation of mean values is seen by the elliptic shape
of the black dashed envelope curve in
Fig.~\ref{fig:AsymmetricGainAndVar}(b) whereas the asymmetric output
variances $\sigma_1^2 \ne \sigma_2^2$ is shown by the black circles
actually being oval shaped.

The asymmetric gain and variances of the input-output map are shown as
a function of $p_{\mathrm{exc}}^{\mathrm{end}}$ in
Figs.~\ref{fig:AsymmetricGainAndVar}(c,d). The different symbols
correspond to those defined in
Fig.~\ref{fig:Influence_of_inversion}(a), and the average gain
$\mathcal{G}$ and variance $\sigma^2$ of these data points were shown
previously in Fig.~\ref{fig:Influence_of_inversion}(b,c) with the
solid blue curves also reproduced. We observe that for inversion
processes of high quality [leading to a small
$p_{\mathrm{exc}}^{\mathrm{end}}$] the asymmetry disappears. Also, if
either of the two inversion pulses is perfect (i.e.~following the
fourth or fifth strategy in Fig.~\ref{fig:Influence_of_inversion}(a)),
the asymmetry also disappears. In
Fig.~\ref{fig:AsymmetricGainAndVar}(c,d) this materializes as the cyan
tip-up triangles and the magenta hexagrams falling on the solid blue
line, which represents the behavior of the average values
$\mathcal{G}$ and $\sigma^2$.

The typical relationship between gain and variance is shown in
Fig.~\ref{fig:AsymmetricGainAndVar}(e). Apart from the ``symmetric
cases'' of cyan tip-up triangles and magenta hexagrams, the data
points seem to follow a general trend. A very low gain parameter and a
gain parameter above unity lead to excess noise. For $\mathcal{G}
\lesssim 1$ there is a regime where one quadrature is squeezed and the
output state has a minor-axis variance of $\sigma_2^2 < \frac{1}{2}$.

Knowing the gain and variance parameters of a linear input-output
relation for $\Xa$ and $\Pa$, it is possible to calculate the fidelity
$F_{\mathrm{q}}$ for a qubit encoded into the $\ket{0}$, $\ket{1}$
Fock states of the cavity field \cite{FidelityPaper}. For the
asymmetric case discussed here, $F_{\mathrm{q}}$ is given by
Eq.~(\ref{eq:Fq}), and the infidelity $1-F_{\mathrm{q}}$ resulting
from all the simulations discussed above is shown in
Fig.~\ref{fig:AsymmetricGainAndVar}(f). In addition, the solid blue
curve is based on the guide-to-the-eye from
Fig.~\ref{fig:Influence_of_inversion}(b,c) using a symmetric formula
for $F_{\mathrm{q}}$ (i.e.~with $\mathcal{G}_1=\mathcal{G}_2 \equiv
\mathcal{G}$ and $\sigma_1^2 = \sigma_2^2 \equiv \sigma^2$). While
Eq.~(\ref{eq:Fq}) presents corrections to such a symmetrized formula
the average parameters $\mathcal{G}$ and $\sigma^2$ govern the
fidelity of the memory protocol quite accurately. The asymmetry does
not in itself present a serious problem for the memory protocol apart
from the fact that the asymmetry only arises when the inversion pulses
are insufficient in strength.

\section{Discussion}
\label{sec:Discussion}
We have investigated the fundamental limitations of a quantum memory
protocol, which uses a spin ensemble for storage and retrieval of a
cavity-field quantum state. The quantum memory performance is fully
characterized by the input-output relation of cavity-field mean values
and variances, parametrized by the mean-value gain $\mathcal{G}$ and
the output state variance $\sigma^2$. In particular, we have examined
how the various parts of the protocol---the storage and read-out, the
waiting times with the spin ensemble decoupled from the cavity, and
the externally driven inversion pulses---affect the obtainable gain
and variance, and the effect of decoherence mechanisms was also
investigated.

We note that the results of the present manuscript, derived for a
single mode of the cavity field, can be extended immediately to the
multi-mode case, as was also shown in
Ref.~\cite{Julsgaard.PhysRevLett.110.250503(2013)}. The only change in
the multi-mode case is a longer $\Tmem$ required to accommodate all the
individual pulses.

Our analysis focused primarily on the impact of an inhomogeneous
spin-frequency distribution. The inhomogeneity arises naturally from
the ensemble nature of the spins and is essential for the multi-mode
capability of the protocol. In some practical realizations,
e.g.~planar wave guide resonators coupled to nitrogen-vacancy centers
in diamond \cite{Kubo.PhysRevLett.107.220501(2011)}, there is also an
inhomogeneous distribution of coupling strengths $g$ for the
spin-cavity interaction. Such an inhomogeneity will in general not
affect our conclusions related to the \emph{linear} regime of the
spin-cavity interaction. However, when inversion pulses are applied,
there may be spin classes which experience insufficient driving while
others are subjected to a nearly perfect inversion process. Such
scenarios can also be examined numerically by our formalism, and we
refer to Ref.~\cite{Julsgaard.PhysRevLett.110.250503(2013)} for a
specific example with inhomogeneous coupling strengths.

\begin{acknowledgments}
  The authors acknowledge support from the EU Seventh Framework
  Programme collaborative project iQIT and the Villum Foundation.
\end{acknowledgments}

\appendix

\section{Special properties of the Lorentzian inhomogeneous
  spin-frequency distribution}
\label{app:Special_Lorentz}
Under the Holstein-Primakoff approximation, and under the assumption
that the initial state of each spin is uncorrelated to its resonance
frequency, the behavior of the Lorentzian broadened spin ensemble
corresponds exactly to that of homogeneous broadening. To show this,
assume in the following $\pauli_z^{(j)} \approx -1$ (the argument also
holds for $\pauli_z^{(j)} \approx 1$) and consider the
Heisenberg-Langevin equations:
\begin{align}
  \frac{\partial\ac}{\partial t} &= -(\kappa+i\Deltacs)\ac
  -i\sum_{j=1}^Ng_j\pauli_-^{(j)} + \hat{f}_a, \\
\label{eq:ddt_pauliMinus_Heisenberg}
  \frac{\partial\pauli_-^{(j)}}{\partial t} &=
     -(\gammaperp+i\Delta_j) \pauli_-^{(j)} - ig_j\ac + \hat{f}_j,
\end{align}
where $\hat{f}_a$ and $\hat{f}_j$ are Langevin-noise operators
accounting for the coupling to the environment (we assume that the
environment coupling experienced by the $j$th spin does not depend on
its resonance frequency $\Delta_j$). A formal integration of the spin
operators leads to:
\begin{align}
    \pauli_-^{(j)}(t) = &\pauli_-^{(j)}(0) e^{-(\gammaperp+i\Delta_j)t}\\
    &+\int_0^t e^{-(\gammaperp - i\Delta_j)(t-t')}[-ig_j\ac(t') + \hat{f}_j(t')]dt'.
\notag
\end{align}
Next, our assumption that $\pauli_-^{(j)}(0)$ is not correlated to
$\Delta_j$ allows for integration over the inhomogeneous ensemble. In
the harmonic-oscillator picture of
Eq.~(\ref{eq:Basic_spin_cavity_evolution}) we find:
\begin{align}
    \b(t) = &\b(0)e^{-(\gammaperp+\frac{w}{2})t}  \\
   &+ \int_0^t e^{-(\gammaperp+\frac{w}{2})(t-t')} [-i\gens \ac(t') + 
    \sum_j \frac{g_j}{\gens}\hat{f}_j] dt', 
\notag
\end{align}
which was derived using the residue theorem being convenient and
applicable for the Lorentzian
distribution~(\ref{eq:f_Lorentz}). Taking the derivative of the above
leads to the coupled equations (with $\hat{f}_b =
\sum_j\frac{g_j}{\gens}\hat{f}_j$):
\begin{align}
\label{eq:ddt_ac_Heisenberg}
  \frac{\partial\ac}{\partial t} &= -(\kappa+i\Deltacs)\ac
     -i\gens \b + \hat{f}_a, \\
\label{eq:ddt_b_Heisenberg}
  \frac{\partial\b}{\partial t} &= 
   -(\gammaperp + \frac{w}{2})\b  - i\gens \ac + \hat{f}_b.
\end{align}
These are identical to the equations for a homogeneously broadened
sample ($\Delta_j = 0$ for all $j$) provided that we replace
$\gammaperp \rightarrow \Gamma = \gammaperp + \frac{w}{2}$.

\section{Gain, variance, and qubit fidelity}
\label{app:GainVarFidelity}
The definition of $\mathcal{G}$ and $\sigma^2$ in
Sec.~\ref{sec:Protocol_general} is generalized in the following. We
note that phase shifts (e.g.~induced by the off-resonant spin-cavity
interaction) may occur in the input-output relations as marked by the
angle $\theta$ in Fig.~\ref{fig:AsymmetricGainAndVar}(a). In addition,
as shown in Fig.~\ref{fig:AsymmetricGainAndVar}(b), the gain and
variance may depend on the phase of the input state since the
inversion pulses may break the symmetry. We cover both of these
scenarios by the input-output mean-value relation:
\begin{equation}
\label{eq:Def_gain_general}
  \begin{bmatrix}
    -X_{\mathrm{c}}^{\mathrm{out}} \\ -P_{\mathrm{c}}^{\mathrm{out}}
  \end{bmatrix}
  = \mat{R}(\theta_1)
  \begin{bmatrix}
    \mathcal{G}_1 & 0 \\ 0 & \mathcal{G}_2
  \end{bmatrix}
  \mat{R}(-\theta_0)
  \begin{bmatrix}
    X_{\mathrm{c}}^{\mathrm{in}} \\ P_{\mathrm{c}}^{\mathrm{in}}
  \end{bmatrix},
\end{equation}
where $\mat{R}(\theta) = \begin{bmatrix} \cos\theta & -\sin\theta \\
  \sin\theta & \cos\theta \end{bmatrix}$ accounts for a
counter-clockwise rotation of angle $\theta$ in the
$(X_{\mathrm{c}},P_{\mathrm{c}})$-coordinate system (corresponding to
multiplying $a_{\mathrm{c}}$ by $e^{i\theta}$). In the symmetric case,
$\mathcal{G}_1 = \mathcal{G}_2 \equiv \mathcal{G}$, the
transformation~(\ref{eq:Def_gain_general}) reduces to a rotation by
the angle $\theta = \theta_1-\theta_0$, being the angle shown in
Fig.~\ref{fig:AsymmetricGainAndVar}(a). For the asymmetric case of
Fig.~\ref{fig:AsymmetricGainAndVar}(b) the angle $\theta =
\theta_1-\theta_0 = 2.3^{\circ}$ accounts for the fact that for input
states located on the $\Xa$- or $\Pa$-axis, the output states are
rotated slightly from the main axes shown as solid lines in
Fig.~\ref{fig:AsymmetricGainAndVar}(b). The angle between the
horizontal $\Xa$-axis and the major axis in this figure is $\theta_1 =
84.2^{\circ}$.

The variance of the output state may depend on the quadrature
phase. Defining $\Xa(\theta) = \Xa\cos\theta + \Pa\sin\theta$, we must
have $\mean{\delta\Xa(\theta)^2} = \mean{\delta\Xa^2}\cos^2\theta +
\mean{\delta\Pa^2}\sin^2\theta + \mean{\Xa\Pa+\Pa\Xa}\frac{\sin
  2\theta}{2}$, and $\mean{\delta\Xa(\theta)^2}^{1/2}$ corresponds to
the angle-dependent radius of the black, oval shapes in
Fig.~\ref{fig:AsymmetricGainAndVar}(b). In all our simulations the
main axes of these oval shapes coincide with the main axes of the mean
value transformation (solid lines in
Fig.~\ref{fig:AsymmetricGainAndVar}(b)), and we define the variance
$\sigma_1^2$ and $\sigma_2^2$ for these main axes. We will generally
define $\mathcal{G}$ and $\sigma^2$ as the average values,
$\mathcal{G} \equiv \frac{1}{2}(\mathcal{G}_1 + \mathcal{G}_2)$ and
$\sigma^2 \equiv \frac{1}{2}(\sigma_2^2 + \sigma_2^2)$, as was done
e.g.~in Fig.~\ref{fig:Influence_of_inversion}(b,c).

For a qubit encoded into the $\ket{0}$ and $\ket{1}$ Fock states of
the cavity field, the quantum memory fidelity is given by:
\begin{equation}
\label{eq:Fq}
  \begin{split}
    F_{\mathrm{q}} &= \frac{1}{6\sqrt{(\sigma_1^2+\frac{1}{2})
     (\sigma_2^2+\frac{1}{2})}}\left\{3 + 
     \frac{3(\sigma_1^2\sigma_2^2-\frac{1}{4})}{(\sigma_1^2+\frac{1}{2})
     (\sigma_2^2+\frac{1}{2})} \right. \\
  &\quad +  \frac{\mathcal{G}_1}{\sigma_1^2+\frac{1}{2}}
  +  \frac{\mathcal{G}_2}{\sigma_2^2+\frac{1}{2}}  
  -\frac{\mathcal{G}_1^2(\sigma_1^2-1)}{(\sigma_1^2+\frac{1}{2})^2}  
   - \frac{\mathcal{G}_2^2(\sigma_2^2-1)}{(\sigma_2^2+\frac{1}{2})^2} \\
  &\quad \left. -\frac{\mathcal{G}_1^2(\sigma_2^2-\frac{1}{2}) 
     + \mathcal{G}_2^2(\sigma_1^2-\frac{1}{2})}
   {2(\sigma_1^2+\frac{1}{2})(\sigma_2^2+\frac{1}{2})}\right\}.
  \end{split}
\end{equation}
The derivation of this formula will be presented elsewhere
\cite{FidelityPaper}.

\section{The cavity-to-spin swapping procedure}
\label{app:cav_spin_swap}
The dynamical evolution of the spin- and cavity-field-mean values
during a resonant transfer of information is discussed mathematically
in this appendix. For a Lorentzian inhomogeneous distribution we
consider the mean values of Eqs.~(\ref{eq:ddt_ac_Heisenberg})
and~(\ref{eq:ddt_b_Heisenberg}) on resonance ($\Deltacs = 0$), and
given an initial cavity field $\mean{\ac(0)} = \alpha$ coupled to the
vacuum spin state $\mean{\b(0)} = 0$, the evolution becomes:
\begin{align}
\label{eq:ac_during_swap}
  \mean{\ac(t)} &= \alpha e^{-\frac{\kappa+\Gamma}{2}t}[\cos(\gens' t) -
   \frac{\kappa-\Gamma}{2\gens'}\sin(\gens' t)], \\
\label{eq:b_during_swap}
  \mean{\b(t)} &= -i\alpha\frac{\gens}{\gens'}  
    e^{-\frac{\kappa+\Gamma}{2}t}\sin(\gens' t),
\end{align}
where $\gens' = \gens\sqrt{1 - \frac{(\kappa-\Gamma)^2}{4\gens^2}}$ is
assumed real to obtain an oscillatory solution. The initial excitation
residing in the cavity is transfered completely to the spin ensemble
at the time $t=\Tswap$ given by:
\begin{equation}
\label{eq:Tswap_analytical}
  \Tswap = \frac{\pi}{2\gens'}\left(1-\frac{2}{\pi}\mathrm{arctan}\left[
    \frac{\kappa-\Gamma}{2\gens'}\right]\right).
\end{equation}
The evolution of the $j$th spin from its initial state
$\mean{\pauli_-^{(j)}(0)} = 0$ is given by formal integration of the
expectation value of Eq.~(\ref{eq:ddt_pauliMinus_Heisenberg}):
$\mean{\pauli_-(t)} = -ig_j\int_0^t
e^{-(\gammaperp+i\Delta_j)(t-t')}\mean{\ac(t')}dt'$. However, the
complexity of Eq.~(\ref{eq:ac_during_swap}) prevents a simple
analytical formula, and we shall consider limiting cases. First, in
the limit $|\Delta_j|\Tswap \gg 1$ the rate of change in
$\mean{\ac(t)}$ is much slower than $\Delta_j$, and by partial
integration we find (neglecting $\gammaperp$ and using
$|\frac{1}{\mean{\ac}}\frac{\partial\mean{\ac}}{\partial t}| \ll
|\Delta_j|$):
\begin{equation}
\label{eq:pauliMinus_Swap_large_Deltaj}
    \mean{\pauli_-^{(j)}(\Tswap)} \approx \frac{g_j e^{-i\Delta_j \Tswap}}{\Delta_j},
\end{equation}
yielding the phase relation of the dotted lines in
Fig.~\ref{fig:SwapIllustration}(c). In the contrary case,
$|\Delta_j|\Tswap \ll 1$, we make a crude approximation of
$\mean{\ac(t)} = \alpha(1-\frac{t}{\Tswap})$ maintaining
$\mean{\ac(0)} = \alpha$ and $\mean{\ac(\Tswap)} = 0$, which leads to
(neglecting $\gammaperp$):
\begin{equation}
\label{eq:pauliMinus_Swap_small_Deltaj}
  \mean{\pauli_-^{(j)}(\Tswap)} \approx -\frac{ig_j\alpha\Tswap}{2}\left(
    1-i\Delta_j\frac{2\Tswap}{3}\right),
\end{equation}
translating into the phase $\phi_j \approx
\frac{\pi}{2}+\frac{2}{3}\Tswap\Delta_j$.

\section{Requirement for numerical simulations}
\label{sec:Requirement_num_sim}
The formal equivalence between Lorentzian and homogeneous broadening
gives rise to analytical expressions for the physical observables
under specific circumstances. For instance, if the initial state at
$t=0$ is the perfectly inverted state, $\mean{\pauli_z^{(j)}} = 1$ and
$\mean{\pauli_x^{(j)}} = \mean{\pauli_y^{(j)}} = 0$, with the cavity
field in the coherent state of amplitude $\mean{\ac} = \alpha$ on
resonance with the spins, $\Deltacs = 0$, the evolution becomes
\cite{Julsgaard.PhysRevA.86.063810(2012)}:
\begin{align}
\label{eq:ac_analytical_decay}
    a_{\mathrm{c}}(t) &= \alpha\frac{(\lambda_++\Gamma)e^{\lambda_+t} -
    (\lambda_-+\Gamma)e^{\lambda_-t}}{\lambda_+ - \lambda_-}, \\
\label{eq:Sminus_analytical_decay}
    S_-^{\mathrm{eff}}(t) &= i\gbar N \alpha \frac{e^{\lambda_+ t} - e^{\lambda_- t}}
      {\lambda_+ - \lambda_-},   
\end{align}
where
\begin{equation}
\label{eq:lambda_num_sim}
  \lambda_{\pm} = -\frac{\kappa+\Gamma}{2} 
    \pm \frac{1}{2}\sqrt{(\kappa-\Gamma)^2 + 4\gens^2},
\end{equation}
and $S_-^{\mathrm{eff}} = \sum_j
\frac{g_j}{\gbar}\pauli_-^{(j)}$. Likewise, for the above perfectly
inverted state and the cavity in vacuum, $\mean{\ac} = 0$, leading to
the initial-state variances, $\mean{\delta\Xa^2} = \mean{\delta\Pa^2}
= \frac{1}{2}$ and $\mean{\delta\S_x^{\mathrm{eff\:2}}} =
\mean{\delta\S_y^{\mathrm{eff\:2}}} = N$, the evolution becomes
(solving the dynamical equations discussed in Sec.~IV A of
Ref.~\cite{Julsgaard.PhysRevA.86.063810(2012)}):
\begin{widetext}
\begin{align}
\label{eq:dXaSqr_analytical_decay}
  \mean{\delta\Xa^2(t)} = \mean{\delta\Xa^2(\infty)} + \frac{\gens^2}
   {(\kappa-\Gamma)^2+4\gens^2}&\left[
   \frac{\Gamma+\lambda_+}{\lambda_+}e^{2\lambda_+ t} + 
   \frac{\Gamma+\lambda_-}{\lambda_-}e^{2\lambda_- t} +
   \frac{\kappa-\Gamma}{\lambda_0}e^{2\lambda_0 t}\right] \\
\label{eq:dSxSqr_analytical_decay}
  \mean{\delta\S_x^{\mathrm{eff}\:2}(t)} = \mean{\delta\S_x^{\mathrm{eff}\:2}(\infty)}
    + \frac{2N \cdot\gens^2}{(\kappa-\Gamma)^2+4\gens^2}&\left[
   \frac{\kappa+\lambda_+}{\lambda_+}e^{2\lambda_+ t} + 
   \frac{\kappa+\lambda_-}{\lambda_-}e^{2\lambda_- t} -
   \frac{\kappa-\Gamma}{\lambda_0}e^{2\lambda_0 t}\right],
\end{align}
\end{widetext}
where $2\lambda_0 = \lambda_+ + \lambda_-$ and identical equations
hold for $\mean{\delta\Pa^2}$ and $\mean{\delta\S_y^2}$. The
steady-state values at $t=\infty$ amount to
\cite{Julsgaard.PhysRevA.86.063810(2012)}:
\begin{align}
\label{eq:CavFieldNoise_SS}
  \mean{\delta\Xa^2} &= \mean{\delta\Pa^2} = \frac{1}{2}\cdot
     \frac{1-C\frac{\kappa-\Gamma}{\kappa+\Gamma}}{1-C}, \\
\label{eq:SpinNoise_SS}
  \mean{\delta\hat{S}_x^{\mathrm{eff}\,2}} &=
    \mean{\delta\hat{S}_y^{\mathrm{eff}\,2}} = N\cdot
    \frac{1+C\frac{\kappa-\Gamma}{\kappa+\Gamma}}{1-C}.
\end{align}
\begin{figure}[t]
  \centering
  \includegraphics{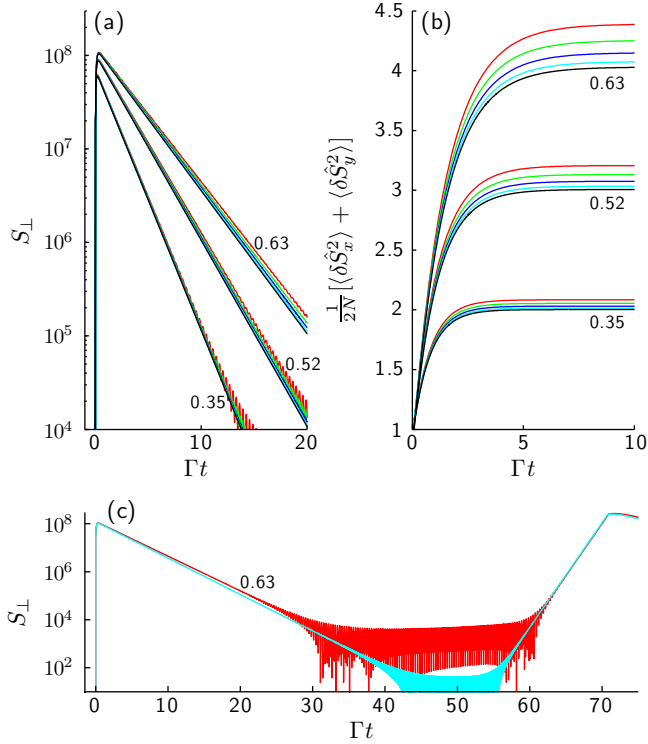}
  \caption{Comparison of numerical simulations with analytical
    results. Panel (a) shows $S_{\perp} = \sqrt{S_x^2+S_y^2}$ for
    three different values of $C$ (indicated on the figure). Within
    each family of curves, from above, the Lorentzian-distribution
    cut-off is varied between values $\Delta_{\mathrm{cut}}/\Gamma =
    20$ (red), 30 (green), 50 (blue), and 100 (cyan), while the lower
    (black) curve is the analytical result calculated from
    Eq.~(\ref{eq:Sminus_analytical_decay}). Panel (b) is arranged
    similar to panel (a), but the normalized spin noise is plotted
    instead. The analytical curves are from
    Eq.~(\ref{eq:dSxSqr_analytical_decay}). Panel (c) extends the
    $C=0.63$ curves from panel (a) for $\Delta_{\mathrm{cut}}/\Gamma =
    20$ (upper red) and 100 (lower cyan) in order to show spurious
    effects from the sub-ensemble discretization. The artificial spin
    revival occurs at $t = \frac{2\pi}{d\Delta} \approx
    71\Gamma^{-1}$.}
  \label{fig:NumericalPrecision}
\end{figure}
We note that these steady-state can be generalized to non-zero values
of $\Deltacs$ by adding the term $[\Deltacs/(\kappa+\Gamma)]^2$ in
both the numerator and denominator. In turn, this leads to the result
of Eq.~(\ref{eq:SpinRef_ConstDet_gain_RESN_mid}).

The above analytical results present an important test base for
numerical simulations. In this article we divide the shape function
$f(\Delta)$ into finite sub-ensembles each of frequency width
$d\Delta$ and subjected to a cut-off $-\Delta_{\mathrm{cut}} \le
\Delta \le \Delta_{\mathrm{cut}}$. In the following we estimate the
numerical effects of the finite $d\Delta$ and $\Delta_{\mathrm{cut}}$.

Numerical simulations of the scenario discussed around
Eqs.~(\ref{eq:ac_analytical_decay})
and~(\ref{eq:Sminus_analytical_decay}) has been shown in
Fig.~\ref{fig:NumericalPrecision}(a) when
$\Delta_{\mathrm{cut}}/\Gamma$ is varied between 20 and 100. Clearly,
since the cut-off limits the frequency bandwidth, the mean-value decay
of the transverse spin component $S_{\perp}$ becomes slower. We also
note that the relative error increases with $C$, but in any case the
simulations converge toward the analytical curve for increasing
cut-off frequencies. For the same values of $\Delta_{\mathrm{cut}}$
the transverse-spin-component variance is shown in
Fig.~\ref{fig:NumericalPrecision}(b) showing a similar trend. The
variance is over-estimated, most pronounced for large values of $C$,
but the correct value is approached for increasing
$\Delta_{\mathrm{cut}}$. The above simulations were performed for
$\gammaperp = 0$, $\gens = 2.5\Gamma$, and number of spins $N =
6.25\times 10^{10}$. The latter determines the quantum-noise limit of
the spin components, $\mean{\S_x^2} = \mean{\S_y^2} = N$, such that
the standard deviation becomes $\sqrt{N} = 2.5\times 10^5$. For
numerical calculations to be faithful, one should ensure that spurious
effects of the sub-ensemble discretization is well below this
limit. In Fig.~\ref{fig:NumericalPrecision}(c) it is exemplified how
an increased $\Delta_{\mathrm{cut}}$ decreases the magnitude of such
effects, and furthermore, that the finite frequency spacing $d\Delta$
causes an artificial revival at $t = 2\pi/d\Delta$. In practical
spin-echo calculations it is required that $d\Delta < 2\pi/\Tmem$ to
avoid these revivals.

\section{Phase shifts induced by a detuned cavity}
\label{sec:stark-shift-effects}
This appendix calculates analytically the impact of a detuned cavity
on the spin ensemble. The initial state is taken as a coherent spin
state, a homogeneous distribution of coupling strengths is assumed, and
the spin-cavity detuning is large, $\Deltacs \gg w,\gens$. The latter
allows the cavity field to be adiabatically eliminated from the
dynamical equations leading to the effective spin Hamiltonian
\cite{Julsgaard.PhysRevA.85.032327(2012)}:
\begin{equation}
  \label{eq:H_cav_eliminated}
  \H = \frac{1}{2}\sum_{j=1}^N\Delta_j \pauli_z^{(j)} - \frac{g^2\Deltacs
   \S_+\S_-}{\kappa^2 + \Deltacs^2}, 
\end{equation}
and the cavity leakage is translated into a correlated spin decay with
$\c = \sqrt{\gammap}\S_-$, $\gammap = \frac{2\kappa g^2}{\kappa^2 +
  \Deltacs^2}$, in the language of the Lindblad part of the master
equation, $\mathcal{L}[\c]\dens = -\frac{1}{2}\cdag\c\dens
-\frac{1}{2}\dens\cdag\c + \c\dens\cdag$. In the Holstein-Primakoff
approximation the spin-ensemble evolution is then governed by:
\begin{equation}
\label{eq:ddt_pauliMinus_elim_cav}
  \frac{\partial\mean{\pauli_-^{(j)}}}{\partial t} = -(\gammaperp + i\Delta_j)
   \mean{\pauli_-^{(j)}} -\frac{ip\zeta}{N}\mean{\S_-},
\end{equation}
where $p = \mean{\pauli_z^{(j)}} = \pm 1$, $N$ is the number of spins,
and $\zeta = \frac{\gens^2(\Deltacs + i\kappa)}
{\kappa^2+\Deltacs^2}$. Now, for an initial coherent spin state all
mean values are equal, $\mean{\pauli_-^{(j)}(0)} \equiv
\mean{\pauli_-(0)}$, and in the spirit of Appendix
\ref{app:Special_Lorentz} the above equation can be integrated over
the inhomogeneous frequency distribution leading to:
$\frac{\partial\mean{\S_-}}{\partial t} = -[\Gamma +
ip\zeta]\mean{\S_-}$, which for a non-inverted ensemble ($p = -1$) has
the solution:
\begin{equation}
\label{eq:Sminus_elim_cav_v1}
  \mean{\S_-(t)} = \mean{\S_-(0)}e^{-\Gamma t}e^{i\zeta t}.
\end{equation}
In turn, by formal integration of
Eq.~(\ref{eq:ddt_pauliMinus_elim_cav}), the evolution of the
individual spins can be deduced:
\begin{equation}
\label{eq:PauliMinus_elim_cav_v1}
  \mean{\pauli_-^{(j)}(t)} = \mean{\pauli_-(0)}e^{-(\gammaperp+i\Delta_j)t}\left[
    1 - \frac{\zeta}{\Delta_j+\frac{iw}{2}+\zeta}\right],
\end{equation}
where it is assumed that the initial mean value, $\mean{\S_-}$, has
vanished due to the frequency inhomogeneity, $t \gg w^{-1}$. The
second term of the effective Hamiltonian~(\ref{eq:H_cav_eliminated})
is only in effect while $\mean{\S_-}$ is not negligible, i.e.~during
the spin-dephasing process, by which the second term in the square
brackets above is acquired. We observe that in addition to the free
evolution, $e^{-i\Delta_j t}$, the spins acquire a complex phase which
varies non-linearly with $\Delta_j$ and hence complicates the
spin-refocusing procedure.

In order to calculate the performance of such a refocusing procedure,
let the spin state evolve until $t = T$, at which time a perfect and
infinitely fast inversion process around the $x$-axis is
performed. Mathematically, this corresponds to taking the conjugate of
the complex spin variables of Eqs.~(\ref{eq:Sminus_elim_cav_v1})
and~(\ref{eq:PauliMinus_elim_cav_v1}) evaluated at the time $T$. At
this time we also allow for changing the cavity parameters such that
$\zeta$ attains a new value, $\zeta'$, for the subsequent
evolution. Repeating the above procedure for the time range $T \le
t\le 2T$ with $p=1$ leads to the ensemble-spin component (assuming $T
\gg w^{-1}$):
\begin{equation}
\label{eq:Sminus_v2_cav_elim_all_orders}
  \mean{\S_-(t)} = \frac{\mean{\S_+(0)}e^{-\gammaperp t}e^{(\frac{w}{2}+i\zeta^*)(t-2T)}}
   {1+\frac{i(\zeta^* + \zeta')}{w}}.
\end{equation}
For completeness, and in order to follow the two-$\pi$-pulse protocol
of Sec.~\ref{sec:spin-cavity-decoupling}, we proceed the evolution of
the state in the time range $2T \le t \le 3T$ maintaining $\zeta'$:
\begin{equation}
\label{eq:Sminus_cav_elim_after_3T}
  \mean{\S_-(t)} = \frac{\mean{\S_+(0)}e^{-\gammaperp t}
   e^{-(\frac{w}{2}+i\zeta')(t-2T)}} {1+\frac{i(\zeta^*+\zeta')}{w}}.
\end{equation}
and after a perfect inversion at $t=3T$ we revert back to the original
$\zeta$ and calculate for $3T \le t \le 4T$:
\begin{equation}
\label{eq:Sminus_cav_elim_after_4T}
  \mean{\S_-(t)} = \frac{\mean{\S_-(0)}e^{-\gammaperp t}
   e^{(\frac{w}{2}-i{\zeta'}^*)(t-4T)}}
   {\left(1-\frac{i(\zeta + {\zeta'}^*)}{w}\right)^2}.
\end{equation}
Evaluated at $t = 4T$ this leads to the predictions of
Eqs.~(\ref{eq:SpinRef_ConstDet_gain})
and~(\ref{eq:SpinRef_ConstDet_theta}).


%

\end{document}